\documentclass{aastex62}

\begin{document}

\def\asec{$^{\prime\prime}$}
\def\lax{{$\mathrel{\hbox{\rlap{\hbox{\lower4pt\hbox{$\sim$}}}\hbox{$<$}}}$}}
\def\gax{{$\mathrel{\hbox{\rlap{\hbox{\lower4pt\hbox{$\sim$}}}\hbox{$>$}}}$}}
\def\farcm{\hbox{$.\mkern-4mu^\prime$}}
\def\farcs{\hbox{$.\mkern-4mu^{\prime\prime}$}}

\def\h{\hskip -4 mm}
\def\ha{\hskip -5 mm}
\def\hb{\hskip -2 mm}
\def\hc{\hskip -1.1 mm}
\def\hd{\hskip -1.6 mm}
\def\hmin{\hskip -0.1 mm}

\title{The Carnegie-Irvine Galaxy Survey. VI. Quantifying Spiral Structure}

\author[0000-0002-3462-4175]{Si-Yue Yu}
\affiliation{Kavli Institute for Astronomy and Astrophysics, Peking University, Beijing 100871, China}
\affiliation{Department of Astronomy, School of Physics, Peking University, Beijing 100871, China}

\author[0000-0001-6947-5846]{Luis C. Ho}
\affiliation{Kavli Institute for Astronomy and Astrophysics, Peking University, Beijing 100871, China}
\affiliation{Department of Astronomy, School of Physics, Peking University, Beijing 100871, China}

\author[0000-0002-3026-0562]{Aaron J. Barth}
\affiliation{Department of Physics and Astronomy, 4129 Frederick Reines Hall, University of California, Irvine, CA, 92697-4575, USA}

\author[0000-0001-5017-7021]{Zhao-Yu Li}
\affiliation{Key Laboratory for Research in Galaxies and Cosmology, Shanghai Astronomical Observatory, Chinese Academy of Science, 80 Nandan Road, Shanghai 200030, China}
\affiliation{College of Astronomy and Space Sciences, University of Chinese Academy of Sciences, 19A Yuquan Road, Beijing 100049, China}
\affiliation{LAMOST Fellow}

\begin{abstract} 
The Carnegie-Irvine Galaxy Survey provides high-quality broad-band optical 
images of a large sample of nearby galaxies for detailed study of their 
structure. To probe the physical nature and possible cosmological evolution 
of spiral arms, a common feature of many disk galaxies, it is important to 
quantify their main characteristics.  We describe robust methods to measure 
the number of arms, their mean strength, length, and pitch angle.  The arm 
strength depends only weakly on the adopted radii over which it is measured, 
and it is stronger in bluer bands than redder bands. The vast majority of clearly 
two-armed (``grand-design'') spiral galaxies have systematically higher 
relative amplitude of the $m=2$ Fourier mode in the main spiral region.  We use
both one-dimensional and two-dimensional Fourier decomposition to measure the 
pitch angle, finding reasonable agreement between these two techniques with a 
scatter of $\sim$2$\degr$. 
To understand the applicability and limitations 
of our methodology to imaging surveys of local and distant galaxies, we create 
mock images with properties resembling observations of local 
($z$ \lax\ 0.1) galaxies by the Sloan Digital Sky Survey and distant galaxies
(0.1 \lax\ $z$ \lax\ 1.1) observed with the {\it Hubble Space Telescope}. These
simulations lay the foundation for forthcoming quantitative statistical 
studies of spiral structure to understand its formation mechanism, 
dependence on galaxy properties, and cosmological evolution.
\end{abstract}


\section{Introduction} \label{sec:intro}

Spiral structure is a fundamental attribute of a large fraction of all disk
galaxies in the nearby Universe, and yet we still lack a full understanding
its nature.  The theory of quasi-stationary density waves proposed 
by \cite{LinShu64, LinShu66} envisaged long-lived spiral arms with constant 
pattern speed. The long-lived spiral arms induce a shock in the gas as it
approaches the bottom of the potential well, compressing it to densities 
conducive to star formation \citep{Roberts1969}. As the newly born stars flow 
downstream from the compression zone, there should be a color gradient across 
spiral arms because of stellar evolution.  However, group transport destroys 
the spiral structure in less than a billion years \citep{Toomre1969}, unless 
the group transport can be slowed down by gas \citep{Ghosh2015}. Later work on 
global mode theory allowed for more steady-state spiral structure 
\citep{Bertin1989a, Bertin1989b, Bertin1996}, predicting that the tightness of the arms 
(their pitch angle) depends on the central mass concentration and the Toomre 
$Q$ parameter of the galactic disk \citep{Bertin1989a, Bertin1989b}.  \cite{Julian1966} 
investigated the response of a particle to an imposed perturbation and found 
that the pitch angle tends to decrease with the shear rate \citep[see also][]{Michikoshi2014}
Another picture is that local instabilities, perturbations, or noise can be 
swing-amplified into dynamical spiral arms \citep{Toomre1981}.  This scenario, 
consistent with results from $N$-body simulations \citep{Carlberg1985, 
Fujii2011, Onghia2013}, predicts that the number of arms is approximately 
inversely proportional to the mass fraction of the galactic disk component.

$N$-body simulations fail to reproduce long-lived grand-design spirals in 
isolated, unbarred disk galaxies without imposing an a priori static spiral 
potential. Instead, spiral arms are transient but recurrent structures 
\citep{Carlberg1985, Bottema2003} that disappear in several 
galactic rotations because of gravitational heating \citep{Sellwood1984}. 
While more recent work shows that the latter effect was overestimated due to 
low numerical resolution and spiral arms can survive much longer in 
high-resolution simulations, the response of the disk to local perturbations is 
still highly non-linear and time-variable \citep{Fujii2011, Onghia2013}.  On 
the other hand, two-armed spirals can be triggered by gravitational 
interaction with a companion galaxy, where the bridge-tail structure induced 
by the tidal force evolves into a symmetric two-armed spiral 
\citep{Toomre1972}. A prototypical example is M51, a grand-design spiral 
galaxy interacting with its companion NGC~5195. The M51 system has been 
reproduced with hydrodynamical simulations by \cite{Dobbs2010}. A number of 
studies focus on the impact on the galaxy from perturbers of different masses. 
A perturber typically needs to be at least 0.01 times the mass of the main 
galaxy to induce a spiral response to the center of the galaxy 
\citep{Byrd1992}.  \cite{Oh2008} confirmed that the strength of the tidally 
induced spiral arms increases with the dimensionless tidal strength but decays
exponentially on a timescale of 1 Gyr. Tidally induced spiral arms can 
persist longer in high-resolution simulations \citep{Struck2011}. If two-armed 
spiral galaxies are indeed mainly driven by the tidal force, one would expect a 
good correlation between spiral arms properties and galaxy environment. 
\cite{Kendall2015} found a strong link between arm strength and the tidal 
force from nearby companion galaxies for their sample of two-armed galaxies. 
Note that the above formation mechanisms of spiral arms are not necessarily 
mutually exclusive.  For example, a tidal interaction theoretically can 
induce an external perturbation, which results in spiral structure obeying the 
density wave theory.

Spiral galaxies are loosely classified into three types according to the 
regularity of their arm structure \citep{Elmegreen1987, Elmegreen1995}: (1) 
grand-design galaxies, with two long arms dominating the galactic disk; (2) 
multiple-armed galaxies, with more than two arms or with two inner symmetric 
arms plus several outer long arms; and (3) flocculent galaxies, with many 
short, chaotic arm segments. Different types of spirals may be linked to 
different formation mechanisms.  Typically, most of the structures in 
flocculent galaxies and the irregular arms in the outer parts of multiple-armed
galaxies are associated with local gravitational instabilities in the gas 
\citep{Goldreich1965} and old stars \citep{Julian1966, Kalnajs1971}.  
Grand-design spiral arms are thought to be tidally induced 
\citep[e.g.,][]{Dobbs2010} and, along with the inner symmetric parts of 
multiple-armed galaxies, may be wave modes obeying density wave theory 
\citep[e.g.,][]{LinShu64, LinShu66, Bertin1996}. However, this kind of 
morphological classification is traditionally based on visual inspection and is 
thus subjective. To make further progress, it is crucial to develop a more 
rigorous, objective method to quantify the key properties of spiral structure,
including the arm number, arm strength, pitch angle, as well as the length of 
arms. 

One-dimensional (1D) and two-dimensional (2D) discrete Fourier decomposition (1DDFT and 2DDFT)
are the two most widely used techniques for analyzing spiral structure 
\citep{ Kalnajs1975, Iye1982, Krakow1982, Elmegreen1989, Puerari1992, Rix1995, 
Block1999, Grosbol2004, Elmegreen2007, Elmegreen2011, Kendall2015}.  
Additional techniques to quantify spiral arm structure have been developed in 
recent years, including a template fitting method \citep{Puerari2014} and  
a method to identify arm segments based on computer vision algorithms
\citep[SpArcFiRe;][]{Davis2014}.
In this work, we apply both of 
1DDFT and 2DDFT to high-quality \emph{BVRI} images of 211 nearby, bright 
southern galaxies selected from the Carnegie-Irvine Galaxy Survey 
\citep[CGS;][]{Ho2011} to characterize their spiral structure.   Using the 
CGS galaxies as a training set, we perform a comprehensive suite of simulations
to evaluate the limits and reliability of our technique when applied to 
large-scale surveys of nearby ($z$ \lax\ 0.1) and distant (0.1 \lax\ $z$ 
\lax\ 1.1) galaxies, using as benchmarks, respectively, the Sloan Digital Sky 
Survey \citep[SDSS;][]{York2000} and the Cosmic Assembly Near-infrared Deep 
Extragalactic Legacy Survey \citep[CANDELS;][]{Grogin2011, Koekemoer2011}. 
A similar image simulation code, FERENGI, has been developed by \cite{Barden2008}, who used SDSS images as input.
Image simulations have also been used to test the limits of nonparametric methods to measure galaxy 
structure \citep{Conselice2003, Conselice2011}, including spiral morphology at high redshift, albeit in a 
limited capacity \citep{Block2001}.
Our image simulations serve as the foundation for forthcoming investigations on the 
statistical properties of spiral structure, their dependence on galaxy 
properties and environment, and possible variations with cosmic epoch.

This paper is organized as follows. Section 2 presents an overview of the 
data used in this study.  Section 3 describes our method to measure the main 
properties of spiral structure.  Section 4 discusses the procedure to generate 
simulated galaxy images to test the limits and uncertainty of our method when 
applied to SDSS and CANDELS survey data.  A summary is given in Section 5.

\section{data} \label{sec:data}
CGS is an optical imaging survey of a statistically complete sample of 605 
bright ($B_T \le 12.9$ mag), nearby (median $D_L=24.9$ Mpc) galaxies in the 
southern sky ($\delta \le 0\degr$). The observations were made using the 
100-inch du~Pont telescope at Las Campanas Observatory in Chile. The parent 
sample comprises $17\%$ elliptical, $18\%$ S0 and S0/a, $64\%$ spiral, and 
$1\%$ irregular galaxies.  The broad-band {\it BVRI} images have a field-of-view 
of 8\farcm9$\times$8\farcm9 and a pixel scale of 0\farcs259.  The median 
seeing of the survey is $\sim 1\arcsec$, and the median surface brightness depth 
reaches 27.5, 26.9, 26.4, and 25.3 mag~arcsec$^{-2}$ in the {\it B}, {\it V}, {\it R}, and 
{\it I} bands, respectively. \cite{Ho2011} describe the survey goals and 
observations, and \cite{Li2011} present the isophotal analysis.  The 
ellipticities and position angles used in this work are mainly taken from 
\cite{Li2011}.

To avoid severe projection effects, which would compromise the study of spiral 
structure, we only choose galaxies with inclination angles $i \leq 60\degr$.
We make use of the star-cleaned images of CGS \citep{Ho2011}, which are free 
from contamination by foreground stars and background galaxies and are thus 
ideal for generating mock images used in our simulation studies of more distant 
galaxies (Section 4).  A minority of galaxies have such serious foreground star 
contamination that even their star-cleaned images cannot be used.  Our final 
sample of 211 disk galaxies spans the full range of Hubble types, whose 
distribution closely follows that of the parent sample (Figure 1).

Warps and spiral arms in the exterior parts of galaxies may distort the outer 
isophotes and thereby introduce uncertainty in the projection parameters, 
ellipticity ($e$) and position angle (PA), measured from the isophotal analysis
(Li et al. 2011).  We consider two other methods to obtain better estimates of 
the projection parameters.  We first assume that the galactic disk is 
intrinsically circular. 
A circular disk becomes elliptical upon projection and thus contains an 
artificial bi-symmetric component. Thus, after performing 
2D Fourier decomposition of the star-cleaned images, the projection parameters 
can be estimated by minimizing the coefficient corresponding to the bi-symmetric
component, $A(m=2,p=0)$ (Equation 9). The radial region occupied by the 
galactic disk was determined by visual inspection to carefully avoid the 
bulge and bar region.  The projection parameters can also be estimated by 
fitting a single or even multiple components to the 2D light distribution of 
the galaxy.  Many CGS galaxies overlap with the {\it Spitzer}\ Survey of 
Stellar Structure in Galaxies \citep[S$^4$G;][]{Sheth2010}. \cite{Salo2015} 
performed human-supervised, multi-component decomposition of S$^4$G galaxies using
{\tt GALFIT} \citep{Peng2010} and derived reliable estimates of $e$ and PA, 
which we use to supplement our work.  The final projection parameters (Table~1)
are used to deproject the galaxies in our selected sample.  If more than one 
set of projection parameters is available, we give preference to that which 
yields a rounder deprojected image or more logarithmic-shaped spiral arms.

\begin{figure}
\figurenum{1}
\centering
\includegraphics[width=10cm]{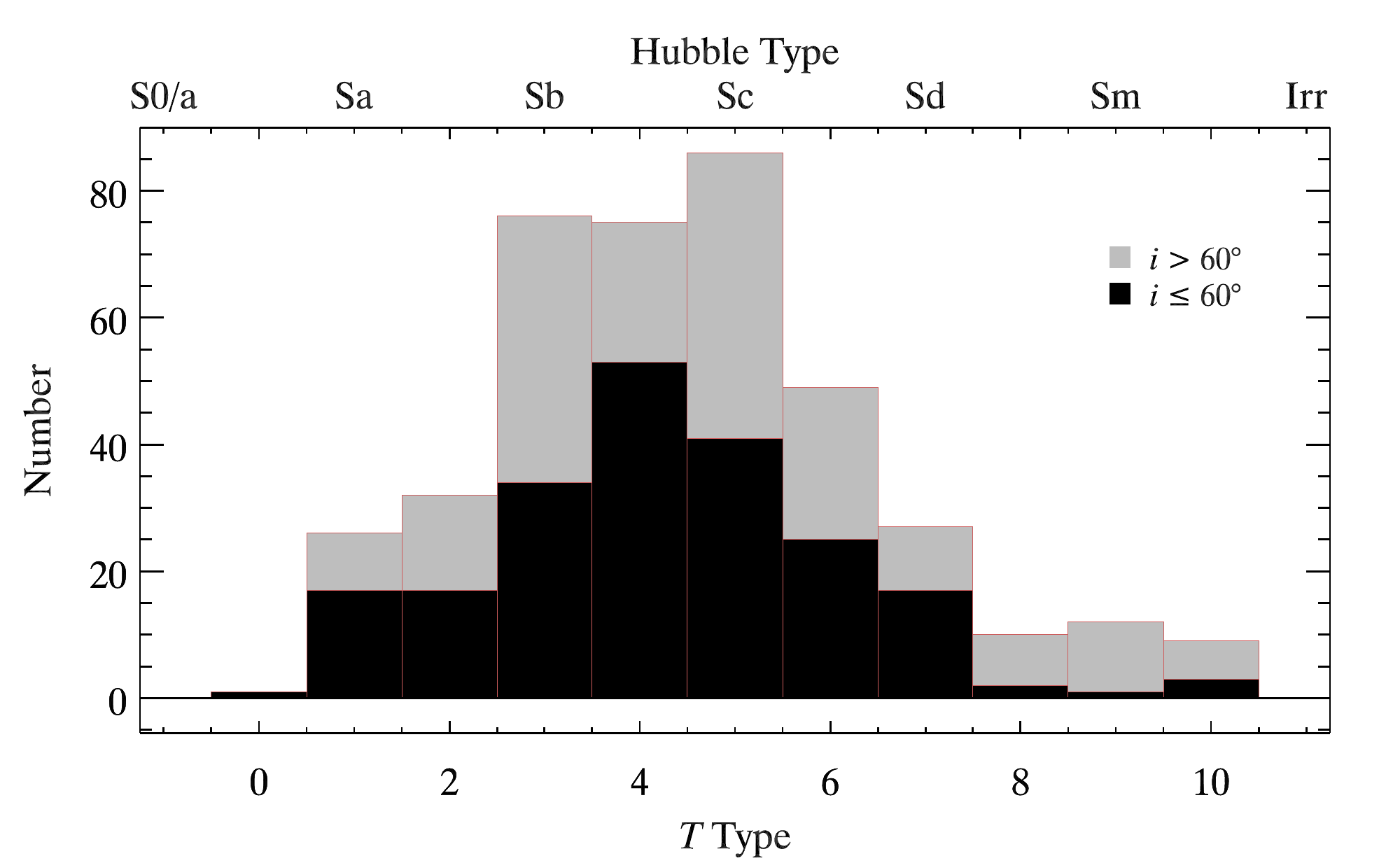}
\caption{Distribution of morphological types for our selected sample and the 
parent sample of CGS spiral galaxies. The bottom axis gives the morphological type index
$T$, with the corresponding Hubble types shown on the top axis. Spiral 
galaxies with inclination angle $i \leq 60\degr$ and $i > 60\degr$ are shown
in black and grey histograms, respectively.}
\end{figure}

\section{Characteristics of spiral arms} \label{sec:method}

\subsection{Strength of Arm Structure} \label{sec:strength} 

1D Fourier decomposition of the azimuthal light profile of 
isophotes can effectively reveal the non-axisymmetric structure present in 
galactic disks.  This technique has been widely used in both observations and 
simulations \citep[e.g.,][]{Elmegreen1989, Elmegreen2007, 
Elmegreen2011, Rix1995, Grosbol2004, Durbala2009, Kendall2015, Baba2015, Hu2016}. The azimuthal 
light profile of isophotes naturally reflects the non-axisymmetric components 
of a galaxy. This method has the advantage that no prior assumption is made 
about the shape of the spiral arm. We study the harmonic components to analyze 
the strength, number, pitch angle, and arm length of spiral structure. The 
star-cleaned images are free from the contamination of foreground stars and 
background galaxies and thus are excellent for studying spiral structure. We 
run the IRAF task {\tt ellipse} on the star-cleaned images with fixed $e$ and 
PA, employing linear steps to extract the azimuthal light profile of the 
isophotes.  Fourier decomposition follows

\begin{eqnarray}
I(r, \theta) = I_{0}(r) + \sum_{m=1}^{6} I_{m}(r) \cos m(\theta + \phi_{m}),
\end{eqnarray}

\noindent
where $I(r, \theta)$ is the azimuthal profile at radius $r$ in direction 
$\theta$, $I_{0}(r)$ is the azimuthally averaged intensity, and $I_{m}(r)$ is the 
cosine amplitude and $\phi_{m}$ the phase angle of each Fourier component.
We perform Fourier decomposition up to $m=6$, because higher order modes 
are nearly negligible. The IDL routine {\tt FFT} was used to calculate the initial 
guesses of the amplitude and phase angle of the Fourier components of each 
profile, and then the {\tt CURVEFIT} routine was used to do the Fourier 
fitting with the initial guesses for $I_{m}(r)$ and $\phi_{m}$.

\begin{figure}[ht!]
\figurenum{2}
\plotone{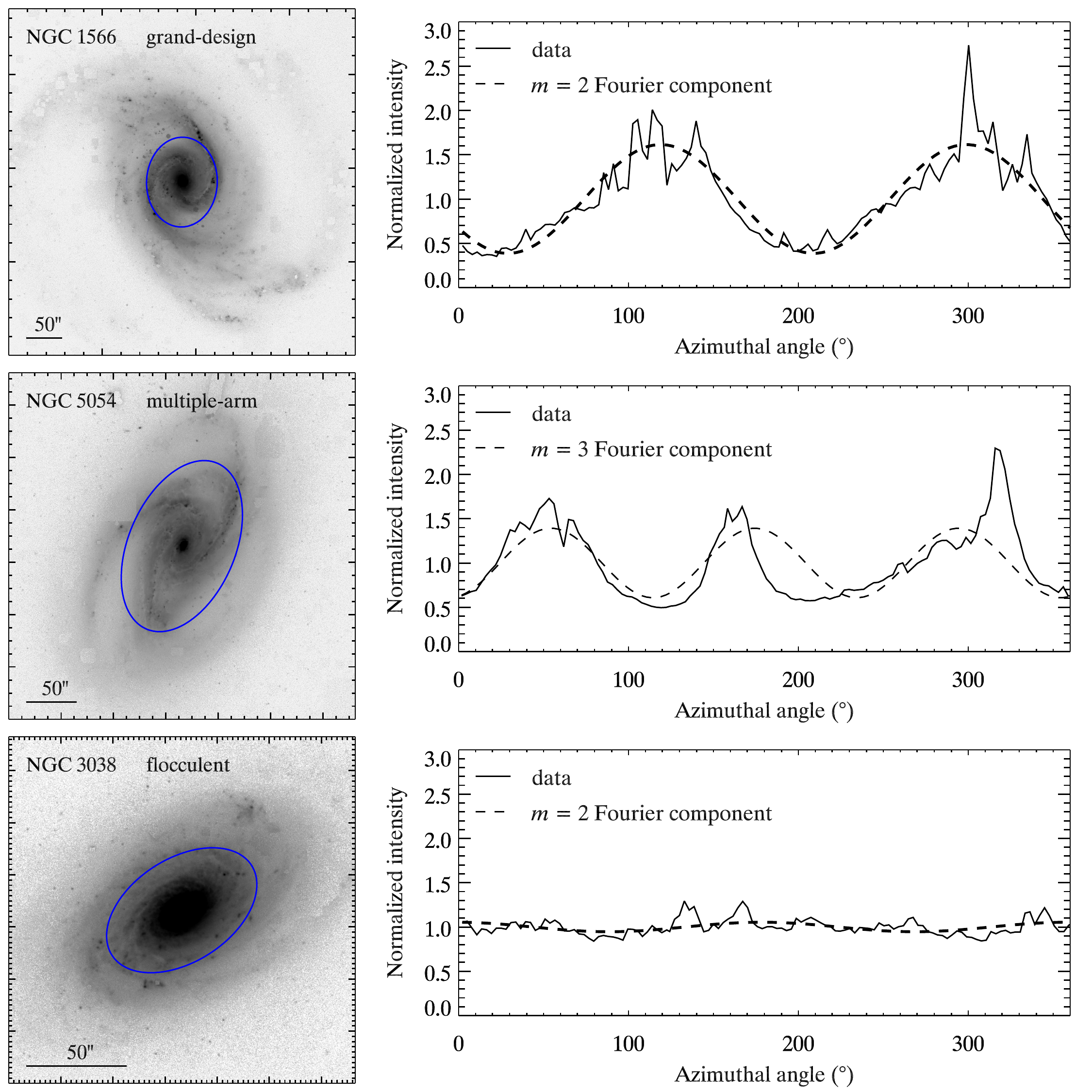}
\caption{Illustration of the three main categories of spiral galaxies: (top 
row) grand-design galaxy NGC~1566, (middle row) multiple-armed galaxy NGC~5054, 
and (bottom row) flocculent galaxy NGC~3038. The images are oriented with north
to the top and east to the left; a scale of 50\asec\ is given in the 
bottom-left corner of each image.  The left column shows the {\it R}-band
CGS image.  The blue ellipse denotes the isophote with which we extract the 
normalized azimuthal intensity profile in the right column.  The dashed line 
represents the best-fit dominant Fourier mode.}
\end{figure}

Figure 2 (right panels) illustrates how the azimuthal light profile of an 
isophote, normalized by its mean intensity (solid line), can effectively 
characterize the relative amplitude of spiral arms. Three categories of spiral 
galaxies are illustrated: NGC~1566, a grand-design galaxy; NGC~5054, a 
multiple-armed galaxy; and NGC~3038, a flocculent galaxy.  For each case, the 
Fourier mode with the highest amplitude is shown as a 
dashed line; the corresponding isophote on the 2D image on the left panels is 
marked by a blue ellipse.  The Fourier mode properly reflects the amplitude 
of the non-axisymmetric structure in the disk, for different types of spiral
arm structure, despite the presence of the sharp and localized features due to bright
H~{\uppercase\expandafter{\romannumeral2} regions along the 
arms \citep[see also][]{Kendall2011}.

The amplitude of the $m$th Fourier component relative to the axisymmetric 
component (the relative amplitude) is defined as

    \begin{eqnarray}
        A_{m}(r) = \frac{I_{m}(r)}{I_0},
    \end{eqnarray}

\noindent
where $m = 1,2,\cdots,6$. Figure 3 illustrates an example of the relative 
amplitudes of the first six Fourier modes as a function of radius for the 
grand-design spiral NGC~1566. 
\begin{figure}
\figurenum{3}
\plotone{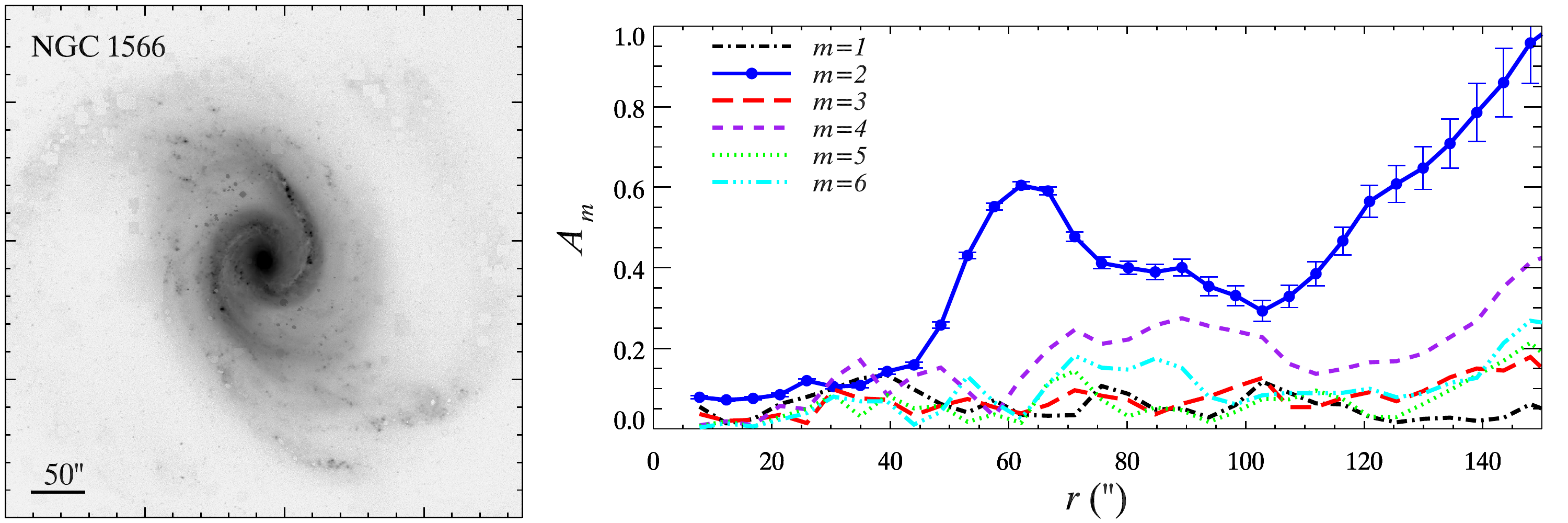}
\caption{Illustration of the Fourier modes of the grand-design spiral NGC~1566.
(Left) {\it R}-band CGS image, oriented with north to the top and east to the left;
a scale of 50\asec\ is given in the bottom-left corner.  (Right) Relative 
amplitudes, $A_m$, of the first six Fourier modes ($m = 1, 2, \cdots, 6$) as a function 
of radius. For the sake of clarity, we only show fitting errors for the $m = 2$ mode.}
\end{figure}
\noindent 
The relative amplitude of the $m=1$ mode ($A_1$) reflects the lopsidedness of 
the galaxy \citep{Rix1995, Bournaud2005, Reichard2008}, while the mean value 
of the $m=2$ mode ($A_2$) is usually used to represent the strength of arm 
structure for two-arm spiral galaxies \citep{Grosbol2004, Durbala2009, Elmegreen2011, 
Kendall2015}. Some multiple-armed and flocculent galaxies, however, 
may have three or even four spiral arms. The $m=2$ mode alone is not enough to 
describe the arm strength for all the different types of spiral galaxies. 
Furthermore, the azimuthal light profile is not perfectly sinusoidal, which 
means that it will contribute not only to the dominant mode but also to other 
Fourier modes.  Therefore, we use the quadratic sum of the $m=2, 3$, and 4 
relative amplitudes to represent the strength of spiral arms at radius $r$:

\begin{eqnarray}
A_{\rm tot}(r) = \sqrt{A_2^2(r)+A_3^2(r)+A_4^2(r)}.
\end{eqnarray}

\begin{figure}[ht!]
\figurenum{4}
\plotone{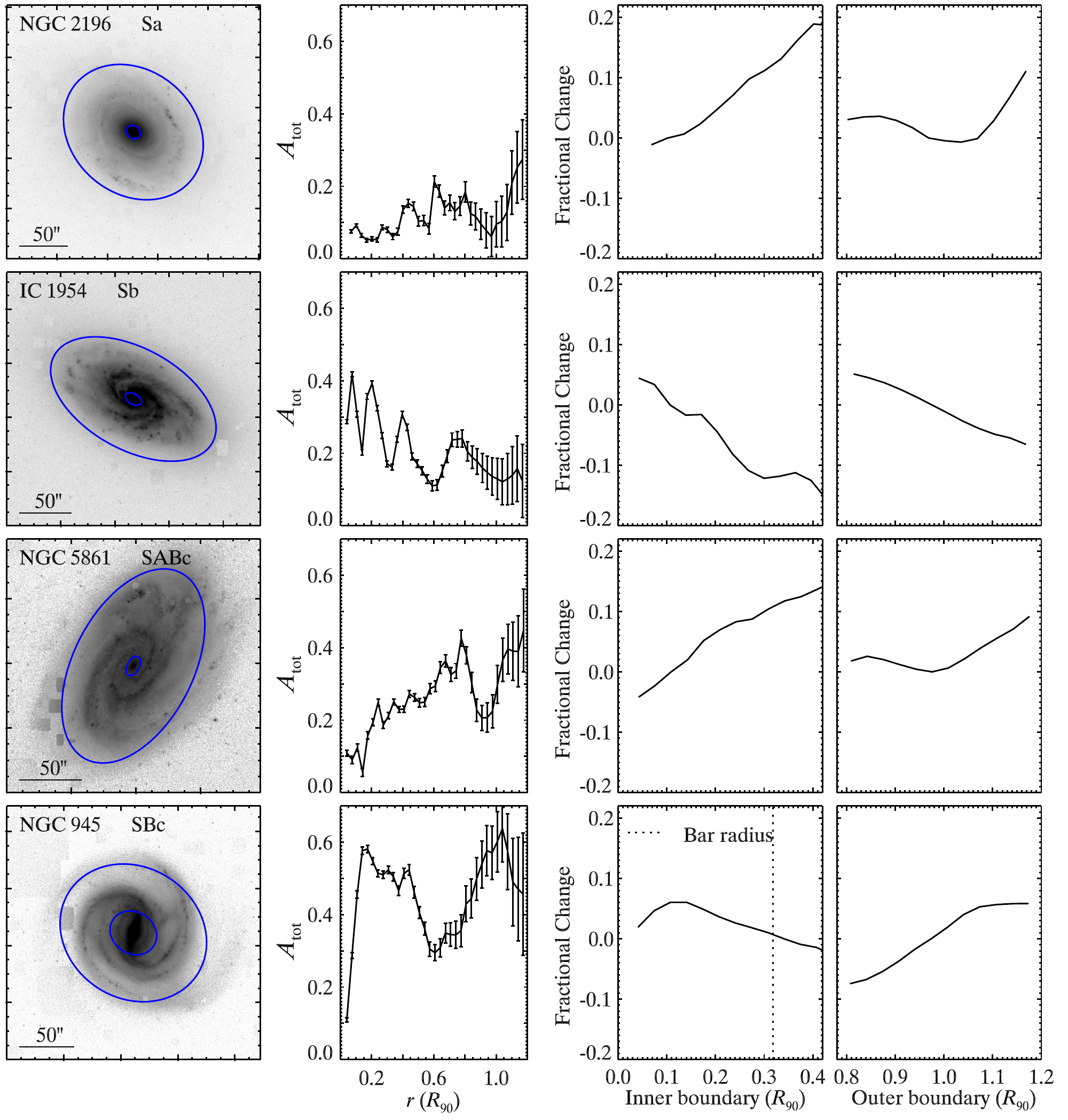}
\caption{The effect of inner and outer boundary radius on the calculation of 
the strength of spiral arms.  The left column shows the {\it R}-band CGS image, 
oriented with north to the top and east to the left; a scale of 50\asec\ is 
given in the bottom-left corner.  The outer blue ellipse marks $R_{90}$,
and the inner blue ellipse marks the radius of the bar, if present, or 
$0.1R_{90}$ if not.  The three columns to the right plot the arm strength 
$A_{\rm tot}$ as a function of radius and the fractional change of the arm 
strength as a function of the inner and outer boundary radius, normalized to 
$R_{90}$.}
\end{figure}

To calculate the mean value of $A_{\rm tot}$, we need to first specify the 
radial range over which the majority of the spiral arm structure is present. The 
spiral arms usually begin at the end of the bar or outside the bulge region 
and terminate at the outskirts of the galaxy. To facilitate the application of 
our method to future studies of galaxies drawn from large-scale imaging surveys,
we require that the boundary of the arm structure should be robustly determined 
and not be very sensitive to observational conditions. Optical studies 
traditionally define the outer boundary of the galactic disk as the radius at 
which the {\it B}-band surface brightness is 25 mag arcsec$^{-2}$.  However, 
$(1+z)^{-3}$ surface brightness dimming and uncertainties in the photometric 
zero point will induce systematic bias or uncertainty for this isophotal 
radius. A definition of radius that is independent of flux normalization, and 
thus independent of redshift and photometric calibration, is needed. The radius
$R_{90}$, which contains 90\% of the light from the galaxy curve of growth, 
naturally meets our requirements.  Our simulated images in Section 4 (outer 
blue ellipse) demonstrate that $R_{90}$ is very robust, varying by less than 
5\% under a variety of observational conditions.  Figure 4 (left column) 
illustrates that $R_{90}$ effectively encompasses the majority of the arm 
structure in galaxies of different types. On the other hand, the inner boundary of the arm 
structure in barred galaxies can be set naturally to the radius of the bar, if present. The bar 
radius is determined based on the large deviation of $e$ and PA near the bar 
region \citep{Li2011}. For unbarred galaxies, we use $0.1R_{90}$ to estimate 
the inner boundary of the arms. Visual inspection reveals that our choice of 
inner boundary (the inner blue ellipse in left column of Figure 4) closely 
delineates the beginning of the arms.  Because the isophotes are extracted 
with fixed $e$ and PA, the presence of a bulge may induce a bimodal 
distribution in the azimuthal light profile, which may lead to an overestimation 
of the relative amplitude of the $m = 2$ mode if the main bulge region is not 
excluded fully interior to $0.1R_{90}$. The third column of Figure 4 shows the 
fractional change of the arm strength when adopting different inner 
boundaries for different galaxy types (Sa to Sc), with the outer boundary 
fixed to $R_{90}$. There is no large systematic overestimation caused by the 
bulge, and the variation caused by the uncertainty in the determination of the 
inner boundary is $< 10\%$.  We also assess the impact of the adopted outer 
boundary. The fractional change of the arm strength (final column of 
Figure 4) is also $< 10\%$ for different choices of outer boundary.

We calculate the mean strength of spiral arms over the radial range from either 
the bar radius or $0.1R_{90}$, depending on whether a not a bar exists, to 
$R_{90}$:

\begin{eqnarray}
	S = \frac{1}{N} \sum_{i=1}^{N} A_{\rm tot}(r_i), 
\end{eqnarray}

\noindent
where $N$ is the number of isophotes within the radial range. The associated 
uncertainty is set as the error of the average value. Similarly, we define the mean 
relative amplitude of the $m$th Fourier component:

\begin{eqnarray}
	S_m = \frac{1}{N} \sum_{i=1}^{N} A_{m}(r_i),
\end{eqnarray}

\noindent
where $m = 1,2, ..., 6$.  

The mean strength of spiral arms may be color-dependent because young, blue 
stars are more concentrated in the arm regions than in the inter-arm regions.
To better understand the wavelength dependence of mean arm strength, we compare
in Figure 5 the relationship between mean arm strength in the {\it I} band ($S_{\it I}$) 
with that in {\it B} ($S_{\it B}$), {\it V} ($S_{\it V}$), and {\it R} ($S_{\it R}$) band.  
It is clear that the 
mean arm strength becomes systematically stronger from the {\it I} to the {\it B} band, 
qualitatively consistent with the notion that spiral arms trigger star 
formation \citep{Roberts1969}.  The correlations between the arm strengths in 
different bands are very tight, with a small total scatter of 0.03.  When using 
the arm strengths for any comparison studies, redshift or bandpass effects can 
be corrected according to the best-fitted linear relations given in the legend. 

\begin{figure}
\figurenum{5}
\plotone{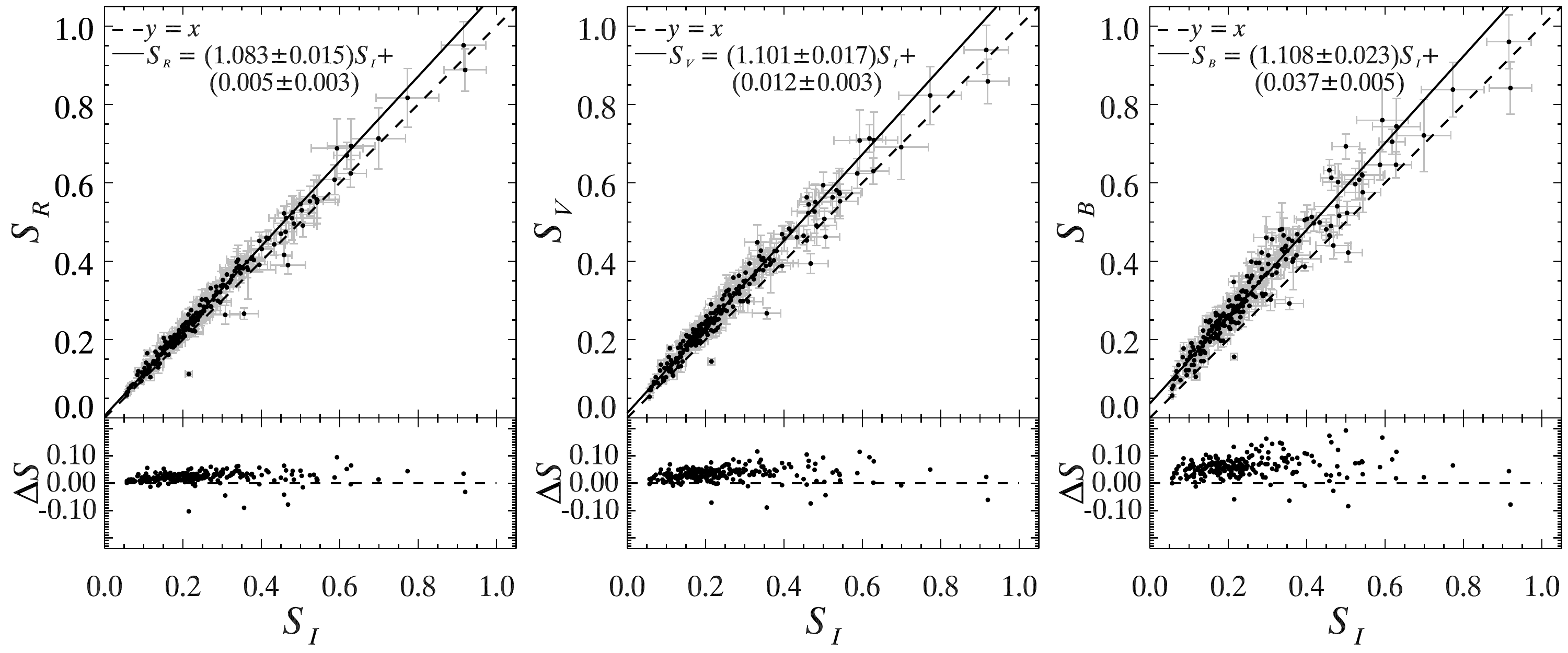}
\caption{Upper row presents
comparison between the mean strength of spiral arms measured in the
{\it I}\ band ($S_{I}$) and that measured in the {\it R} ($S_{R}$, left), {\it V} ($S_{V}$, middle), and 
{\it B}\ ($S_{B}$, right) band. 
Bottom row presents the difference of the mean strength, $\Delta S=S_{R,V, \text{or} B} - S_{I}$, 
as a function of $S_I$. 
The solid  line marks the best-fit linear relation,
while the dashed line denotes $y=x$ and $\Delta S=0$ in the upper panels and bottom panels, respectively.
}
\end{figure}

\subsection{Pitch Angle and Length of Spiral Arm from 1D Method} \label{sec:armlength} 

Grand-design galaxies have two symmetric, long spiral arms. Some multiple-armed 
and flocculent galaxies can also have inner symmetric arms, but they split up 
into more fragmentary pieces at larger radii.  Symmetric arm patterns may be 
wave modes obeying density wave theory.  To quantitatively identify and study 
grand-design arms, we need to measure their pitch angle and length.  Some 
studies \citep{Grosbol2004, Kendall2011} use 1D Fourier decomposition to 
calculate the pitch angle.  The phase of a spiral wave can be identified by 
the phase angle of the $m = 2, 3,$ or $4$ Fourier mode, and the variation of 
the phase angle with radius can reveal useful information, such as the pitch 
angle of the spiral arm.  

Our 1D method is based on the star-cleaned images and the phase angle from Equation (1).
We begin by defining the region occupied by symmetric spiral arms as the 
``main'' spiral region, within which the phase angle $\phi$ of the Fourier 
mode corresponding to the spiral pattern continuously increases or decreases
with radius. The slope of the phase angle profile as a function of radius in 
this main spiral region reflects the tightness of the spiral arms. If a bar is
present, the phase angle will remain almost constant before the main spiral 
region, then abruptly changes at the transition between the bar and arms; in the 
absence of a bar, the phase angle profile usually shows no regularity and can 
be identified easily.  Beyond the main spiral region, the phase angle no 
longer changes monotonically.  It can even become chaotic because the continuous 
arm structure terminates or transforms into feathery structures dominated by
higher-frequency modes. We choose the Fourier mode that exhibits the most 
regular phase angle profile, showing the most nearly monotonic variation with radius, as the 
representative Fourier mode, $M_{\text{1D}}$, of the spiral 
pattern. We determine the main spiral region based on the behavior of the phase 
angle profile of Fourier mode $m=M_{\text{1D}}$. 
Since the cosine function in Equation (1) is symmetric, what we identify is the symmetric
part of the spiral pattern.
Assuming that the arms are logarithmic, the 
$\phi-r$ diagram of the main spiral region is fitted with a logarithmic function

		\begin{eqnarray}
			\phi = b \cdot {\rm ln}\, r + {\rm const.},
		\end{eqnarray}

\noindent
where $\phi$ is the phase angle, $r$ is the radial distance from the 
center, and $b$ is a coefficient.  The pitch angle $\varphi$ is given by

		\begin{eqnarray}
			\varphi = \text{arctan}\left( \frac{1}{b} \right),
		\end{eqnarray}
with its uncertainty determined through propagation of the fitting error of $b$.
The length of the symmetric spiral pattern can be estimated from

\begin{eqnarray}
L_{\rm arm} = \sqrt{1+b^2} \cdot \left(r_{\rm max} - r_{\rm min} \right),
\end{eqnarray}

\noindent
where the $r_{\rm min}$ and $r_{\rm max}$ are the minimum and maximum radial 
extent of the main spiral region.
The uncertainty of $L_{\rm arm}$ is determined through propagation of the fitting error of $b$.
	
Figure 6 shows an application of the method to measure the pitch angle and 
arm length for NGC~1566. 
The blue data points in the right panel denote the 
$m = 2$ phase angle profile ($\phi_{\text{m}=2}$), and their corresponding positions on the 
spiral arms are shown in the left panel. The angle $\phi_{\text{m}=2}$ monotonically increases within the main
spiral region $(30\arcsec,130\arcsec)$, over which the pitch angle and arm length can be calculated.
The best-fit logarithmic function, marked by the solid red curve,  gives a pitch angle of 
$19\fdg5\pm0\fdg5$ and arm length of $298\arcsec\pm7\arcsec$.
A slight variation ($\sim10\%$) of endpoints of the main spiral region will give rise to an uncertainty of
$\sim0\fdg5$ and $\sim6\arcsec$ for pitch angle and arm length, respectively, which are comparable with the
uncertainties from the Fourier fitting.

A strong, underlying assumption of the above method is that the arm structure 
is dominated by a single ($m$th) Fourier mode. The method fails if the arm 
structure of the galaxy is very flocculent or if it is not symmetric at all, in 
which case there will be no main spiral region or continuously changing 
phase angle profile.  Galaxies whose spiral arms are only marginally symmetric
will have their symmetric part identified.  Another shortcoming of this method
is that it cannot identify the main spiral region if an arm runs almost 
parallel to the ellipse, as can occur in some strongly barred galaxies. 

Table~1 lists $M_{\rm 1D}$, the radial range of the main spiral region, pitch 
angles, and arm lengths for 168 galaxies measured using the 
1D method.  We also give $S_{2,\text{main}}$, the mean relative amplitude of the 
$m = 2$ mode in the main spiral region. We compute this quantity regardless of 
whether there are two arms or not.  We expect $S_{2,\text{main}}$ to be 
stronger in grand-design galaxies than in non-grand-design galaxies.  As 
demonstrated in Section 3.4, $S_{2,\text{main}}$ is very useful and effective 
to distinguish grand-design galaxies from other types, and thus can be used to 
probe the formation mechanism of spiral arms.

\begin{figure}
\figurenum{6}
\plotone{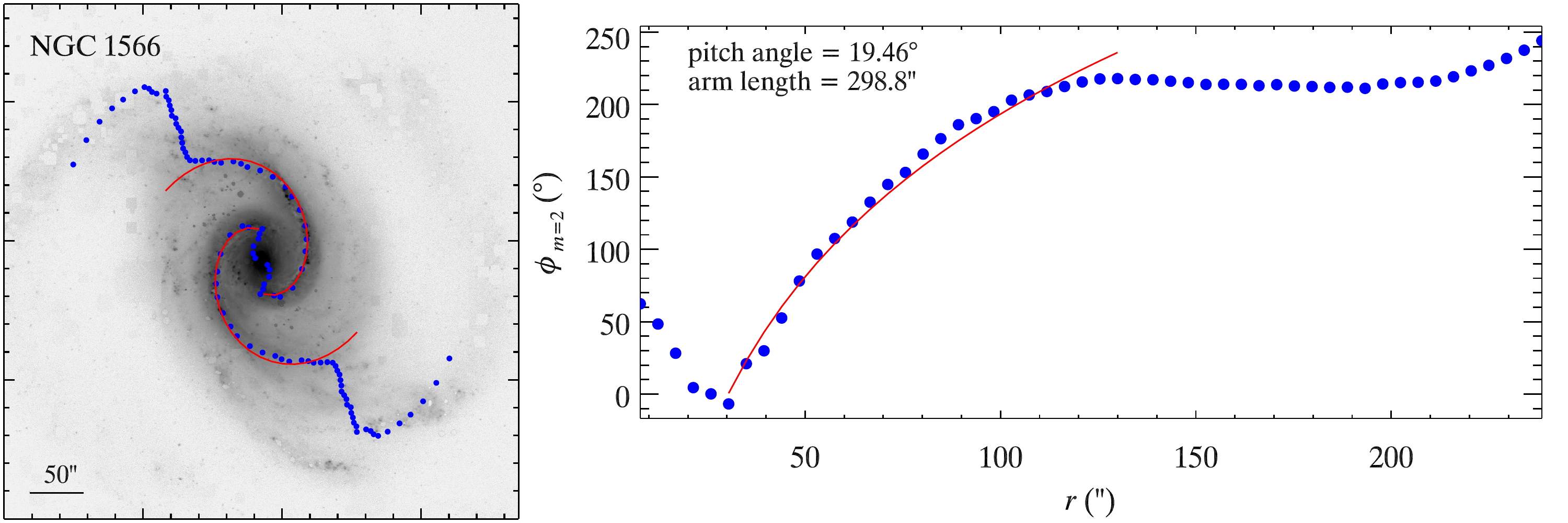}
\caption{Illustration of our 1D method to measure the pitch angle of spiral 
arms.  (Left) {\it R}-band CGS image of NGC~1566, oriented with north to the top 
and east to the left; a scale of 50\asec\ is given in the bottom-left corner.  
The blue points mark the positions of the peak of the $m=2$ Fourier component, 
and the red line traces the best-fit logarithmic function.  (Right) The $m=2$ 
phase angle profile (blue points) and the best-fit logarithmic function (red 
line). The derived pitch angle is 19$\fdg$46, and the arm length is 298\farcs8.}
\end{figure}

\subsection{Pitch Angle from 2D Method} \label{sec:pitch}    

Another, more widely used technique to measure pitch angle is 2DDFT 
\citep[][]{Kalnajs1975, Iye1982, Krakow1982, Puerari1992, Puerari1993, 
Block1999, Seigar2005, Davis2012}. The resulting pitch angle measurements from 
1DDFT and 2DDFT are not necessarily the same, because the basic principle is 
different: the 1D method tries to identify the position of spiral arms and 
then fits a logarithmic function to those positions; the 2D method selects a 2D 
Fourier mode to represent the spiral arm. An important technical 
question arises:  do these two methods give consistent results? In order to 
address this question and better understand the uncertainty induced by the 
different techniques, we also develop a 2D method based on 2DDFT to measure 
pitch angles.

We begin by deprojecting the galaxy image to face-on orientation. The sky 
background is first subtracted from the star-cleaned image. Then, with the PA
and axis ratio ($1 - e$) well defined, the galaxy is deprojected by rotating 
the sky-subtracted, star-cleaned image to align the major axis of the disk 
in the vertical direction and 
then using the IRAF routine {\tt geotran} to stretch the x-axis by the axis ratio.
This process assumes that the disk of the galaxy is intrinsically circular.  
To apply 2DDFT, the deprojected, sky-subtracted, star-cleaned galaxy image in 
Cartesian coordinates is transformed into an image in polar coordinates, and 
then the light distribution is decomposed into a superposition of sinusoidal 
functions of different frequency, with corresponding amplitude given by

\begin{eqnarray}
A(m, p) = \frac{1}{D} \int ^{\text{ln}(r_{\text{out}})}_{\text{ln}(r_{\text{in}})} \int^{\pi}_{-\pi} \sum_{j=1}^{N} I_j(r_j, \theta_j) \delta(\mu-\mu_j) \delta(\theta-\theta_j) e^{-i(m\theta + p \mu)} d\theta d\mu,
\end{eqnarray}

\noindent
with $D$ a normalization factor expressed by

\begin{eqnarray}
D =  \sum_{j=1}^N I_j,
\end{eqnarray}

\noindent
where $I_j$ is the intensity of the $j$th pixel at $(r_j, \theta_j)$, 
$r_{\text{in}}$ and $r_{\text{out}}$ are the inner and outer radius of the 
spiral structure, respectively, $N$ is the number of pixels within the radial 
range, and $\mu \equiv {\rm ln}\, r$. 
For barred spiral galaxies, the spiral 
arms usually begin at the end of the bar. Although the bar lengths of CGS galaxies were
provided by \cite{Li2011} using isophotal analysis, unfortunately 
these measured bar lengths do not always fully exclude the bar. Underestimation
of bar length causes large overestimation of the pitch angle, if the bar length is adopted to represent the
inner boundary of spiral arms. \cite{Davis2012} proposed to determine the inner 
radius by assuming the pitch angle is stable beyond the bar. However, the 
pitch angle can change with radius, by as much as 20\% \citep{Savchenko2013}.
To better determined the inner boundary of spiral arms, we first generate an
unsharp-masked image by dividing the star-cleaned image by its
blurred version, using, as kernel, a Gaussian function with full width at half 
maximum (FWHM) $0.1 r_{\rm out}$, 
which is large enough to smooth large-scale structures such as a bar or spiral
arms. This procedure effectively highlights the bar and spiral structure. Then we 
manually determine the inner radius
after carefully avoiding the bi-symmetric component (bar or lens), which 
may induce  a central peak in the resulting power spectrum and cause measurement errors.
The outer radius is set to the radius where the spiral arms disappear, but 
2DDFT is not sensitive to its adopted value.

Figure 7 presents the Fourier spectra (for radial range 
$[r_{\rm in}, r_{\rm out}]$) for the deprojected images of NGC~1566 
and IC~4538.  In most cases, the resulting pitch angle is the pitch angle of 
the most prominent peak in the Fourier spectra.  The grand-design galaxy 
NGC~1566 shows a dominant $m = 2$ mode in its spectra, and thus $p$, 
corresponding to the highest amplitude ($|A(m, p)|$) indicated by the arrow, 
is used to calculate the pitch angle

\begin{eqnarray}
\varphi = \arctan{\left(- \frac{m}{p} \right)}.
\end{eqnarray}

\noindent
The upper-left panel of Figure 7 shows the deprojected image of NGC~1566 
overplotted with a synthetic arm with a pitch angle of 21$\fdg$13
(its orientation is adjusted manually)}, illustrating 
that 2DDFT can accurately derive pitch angles for galaxies with clear and 
symmetric two-arm spirals. The resulting pitch angle is consistent with that 
obtained from 1DDFT (19$\fdg$46; Figure 6). However, for some multiple-armed 
or flocculent galaxies, the disk structure is so complicated that the highest 
amplitude does not contribute to the spiral pattern. The bottom-right panel of 
Figure 7 shows an example of Fourier spectra for this kind of galaxies. 
The highest amplitude of IC~4538 may correspond to some other bi-symmetric 
component, although it is inconspicuous in the deprojected image, or it may be just 
caused by the complicated spiral structure. If the most prominent peak were 
chosen to do the calculation, it would lead to a severely incorrect (usually overestimated) pitch angle.
Instead, we identify the 
secondary maxima, the highest amplitude 
of the $m = 4$ Fourier mode (purple arrow in Figure 7), to calculate a 
pitch angle of 19$\fdg$98, which produces synthetic arms that match well the 
deprojected image and is consistent with the value of 20$\fdg$0 derived from
the 1D method. The 2D Fourier mode identified for further calculation is denoted
as $M_{\rm 2D}$ in Table~1.

\begin{figure}
\figurenum{7}
\plotone{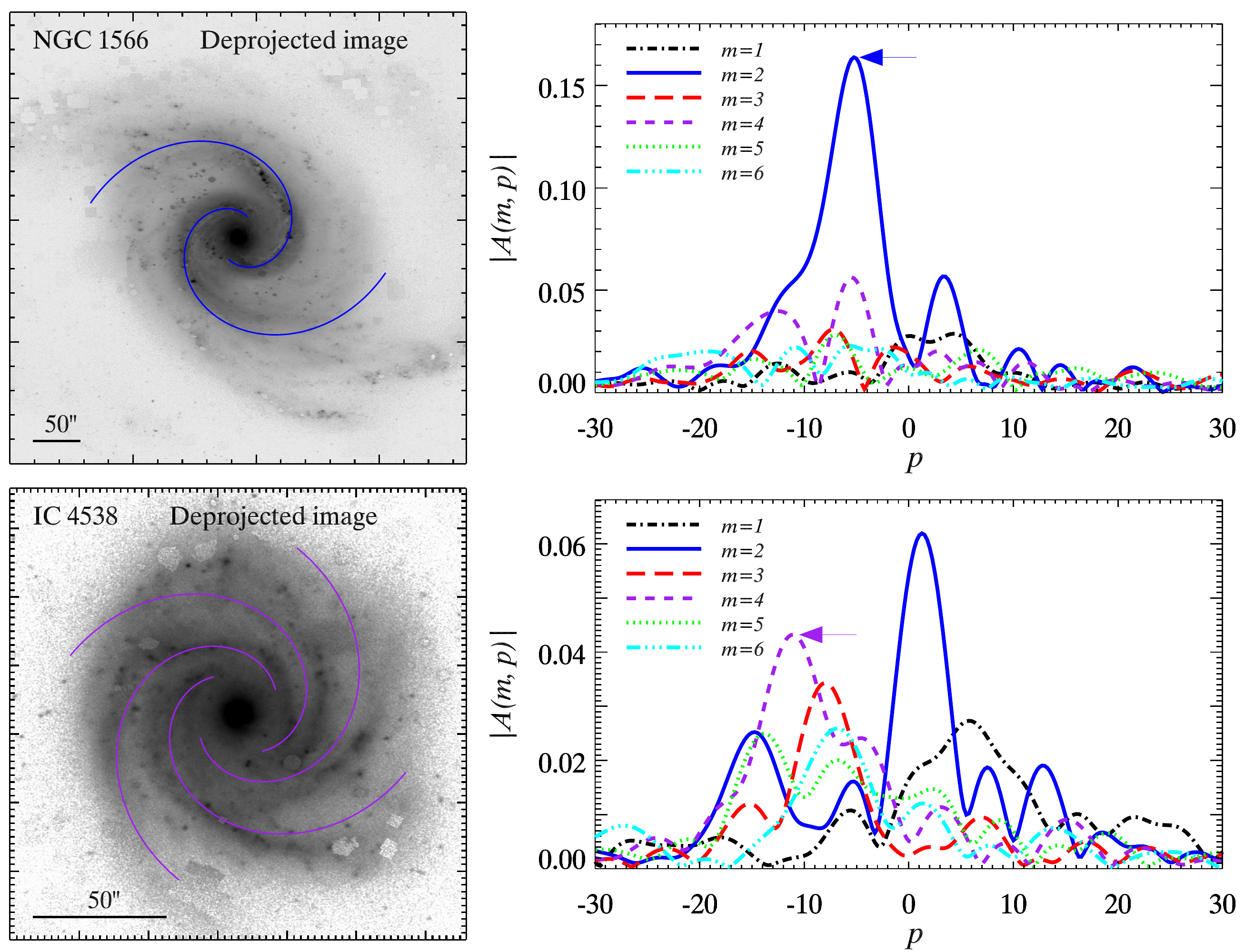}
\caption{Illustration of our 2D method to measure the pitch angle of spiral 
arms.  (Left) Deprojected {\it R}-band CGS images of NGC~1566 and IC~4538, 
oriented with north to the top and east to the left; a scale of 50\asec\ is 
given in the bottom-left corner.  (Right) Amplitude of the 2D spectra for the 
first six Fourier modes.  The pitch angle is calculated from the peak of the 
$m=2$ mode in NGC~1566 and from the peak of the $m=4$ mode in IC~4538.  The 
solid lines superposed on the deprojected images mark the synthetic arms with 
the resulting pitch angle.}
\end{figure}

If the arm structure is too irregular or too faint to be easily recognized, the
Fourier spectra show large variation with no prominent peak, and the pitch 
angle is unmeasurable.
The 2D method also fails if the arms deviate drastically from logarithmic shape, 
which may occur in galaxies with very long bars. We visually check the 
deprojected images of every galaxy overplotted with the synthetic arm pattern 
to verify the fidelity of the derived pitch angles.

2DDFT requires no prior knowledge of the light distribution. However, once 
we select the maxima amplitude of a particular Fourier mode $m$ to calculate 
the pitch angle, we do assume that the spiral structure can be represented by 
the symmetric, logarithmic $m$ Fourier component. 
Even though the spiral arms are not perfectly symmetric, as in the case of IC~4538 shown 
in Figure 7 (bottom-left panel), the 2D method can still correctly measure the pitch angle.
If the arms deviate from a 
logarithmic function, the pitch angle changes with radius. We quantify the 
uncertainty introduced by this effect and that associated with the
determination of the radial extent of the spiral arms
by performing the 2DDFT with window 
functions. In addition to measuring pitch angle with radial bins 
$(r_{\text{in}}, r_{\text{out}})$, we also repeat the calculation for another 
three radial bins: $(r_{\text{in}}, r_{\text{out}}-0.2\Delta r)$, 
$(r_{\text{in}}+0.2\Delta r, r_{\text{out}})$, and 
$(r_{\text{in}}+0.1\Delta r, r_{\text{out}}-0.1\Delta r)$, where 
$\Delta r \equiv r_{\text{out}}-r_{\text{in}}$. Our final value of the 2D 
pitch angle is the average of these four estimates, with the uncertainty given
by their standard deviation. The uncertainty due to the resolution of $p$ in Equation (9), 
which is set as 0.25 in this work, is negligible. Table~1 lists $M_{\rm 2D}$, radial range, and 
pitch angle measured using the 2D method for 159 galaxies.
 
We use both 1DDFT and 2DDFT to estimate the pitch angle of spiral 
arms. For galaxies with a continuous and symmetric arm pattern, like that of 
NGC~1566, both methods give consistent results.  Galaxies with asymmetric 
or irregular arms, however, may produce different results for the two methods. 
Figure 8 gives a direct comparison between the pitch angles derived from both 
the 1D and 2D methods.  A small, systematic trend may be present.
For pitch angles $\lesssim 20\degr$, the 1D method tends to yield somewhat 
larger pitch angles than the 2D method, whereas for pitch angles $\gtrsim 
20\degr$ the opposite seems to hold.  However, the scatter between the two 
sets of measurements is only $\sim 2\degr$, indicating that overall both 
techniques produce essentially consistent results.

The number of spiral arms could be set as the strongest 
1D or 2D Fourier mode. Unfortunately, this cannot correctly 
describe the number of spiral arms. For example, IC~4538 clearly has four arms, but the strongest
Fourier mode is $m=2$, both in 1D and 2D.  The number of arms is taken to be 
the Fourier mode chosen to calculate the pitch angle: $M_{\text{1D}}$ and $M_{\text{2D}}$. They
are not necessary the same, especially for some multiple-armed galaxies whose arms change from an 
inner two-arm structure into multiple arms in the outer region. The number of spiral arms is not well-defined 
in such cases. Therefore, we take $M_{\text{1D}}$ and $M_{\text{2D}}$ as the range of the number of
spiral arms.  The spirals of a few galaxies are strongly distorted by a long bar (e.g., NGC~1300); even 
though such galaxies have two arms, their pitch angle
is unmeasurable and neither $M_{\text{1D}}$ nor $M_{\text{2D}}$ is available.

Figure 9 (upper row) plots the pitch angle against Hubble type. We confirm 
that the spiral arms on average tend to be more tightly wound in galaxies with earlier Hubble type, 
but with large scatter in pitch angle ($\sim8\degr$) for a given Hubble type. 
A dependence of pitch angle on Hubble type is expected in the context of density wave 
theory \citep{LinShu64, Roberts1975, Bertin1989a, Bertin1989b}.
Our results are consistent 
with those of \cite{Kennicutt1981}, \cite{Ma2002}, and \cite{Kendall2015}; no evidence of such correlation was
found by \cite{Seigar1998}, which is partially due to their small range of measured pitch angle.
There is no obvious correlation between pitch angle and Elmegreen arm class (bottom row of Figure~9). 

\begin{figure}
\figurenum{8}
\plotone{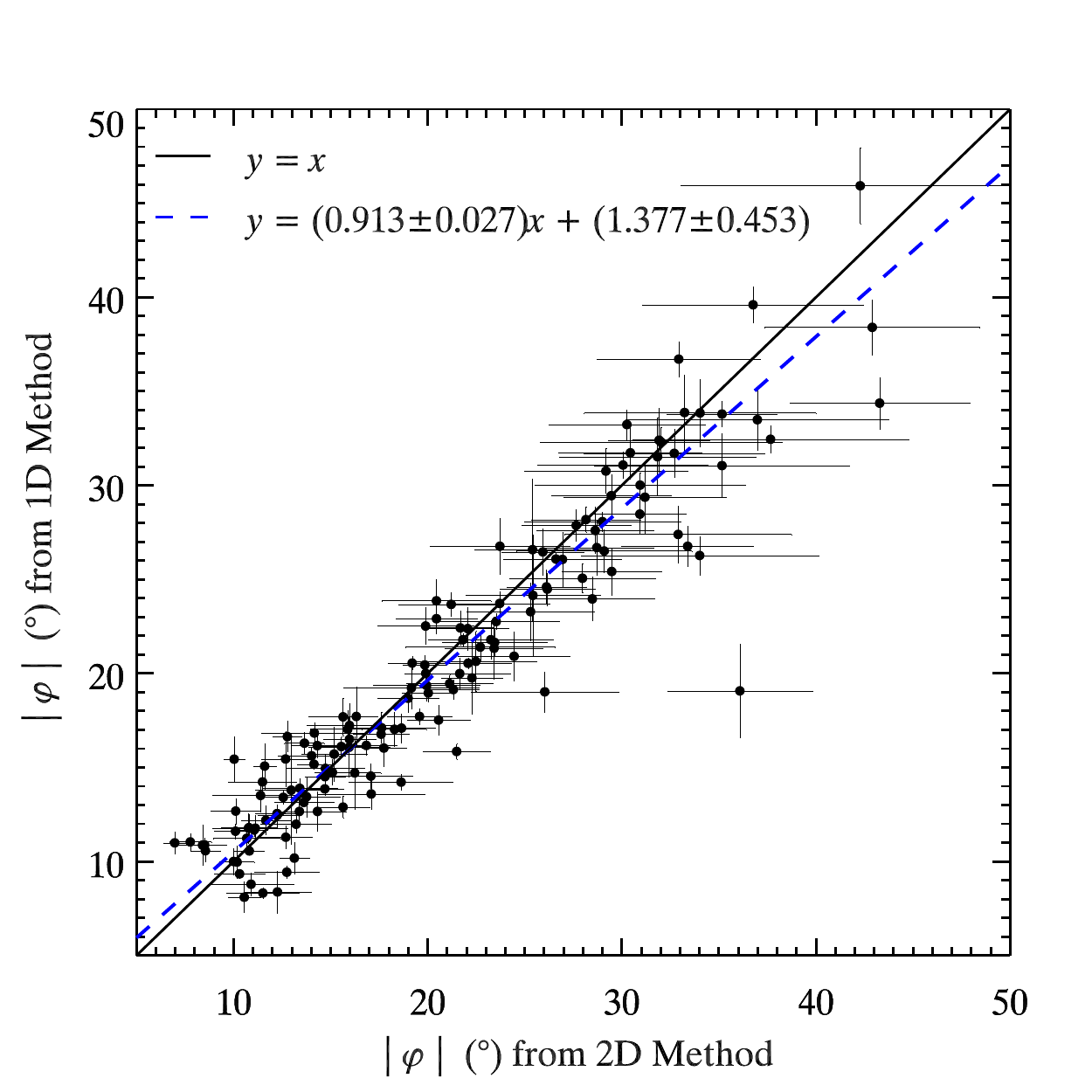}
\caption{Comparison of pitch angles for 146 spiral galaxies, obtained from
both the 1D and 2D methods. The black solid line denotes $y=x$, while the 
solid dashed line is the best-fit linear relation. The correlation has a scatter 
of $\sim 2\degr$.}
\end{figure}

\begin{figure}
\figurenum{9}
\plotone{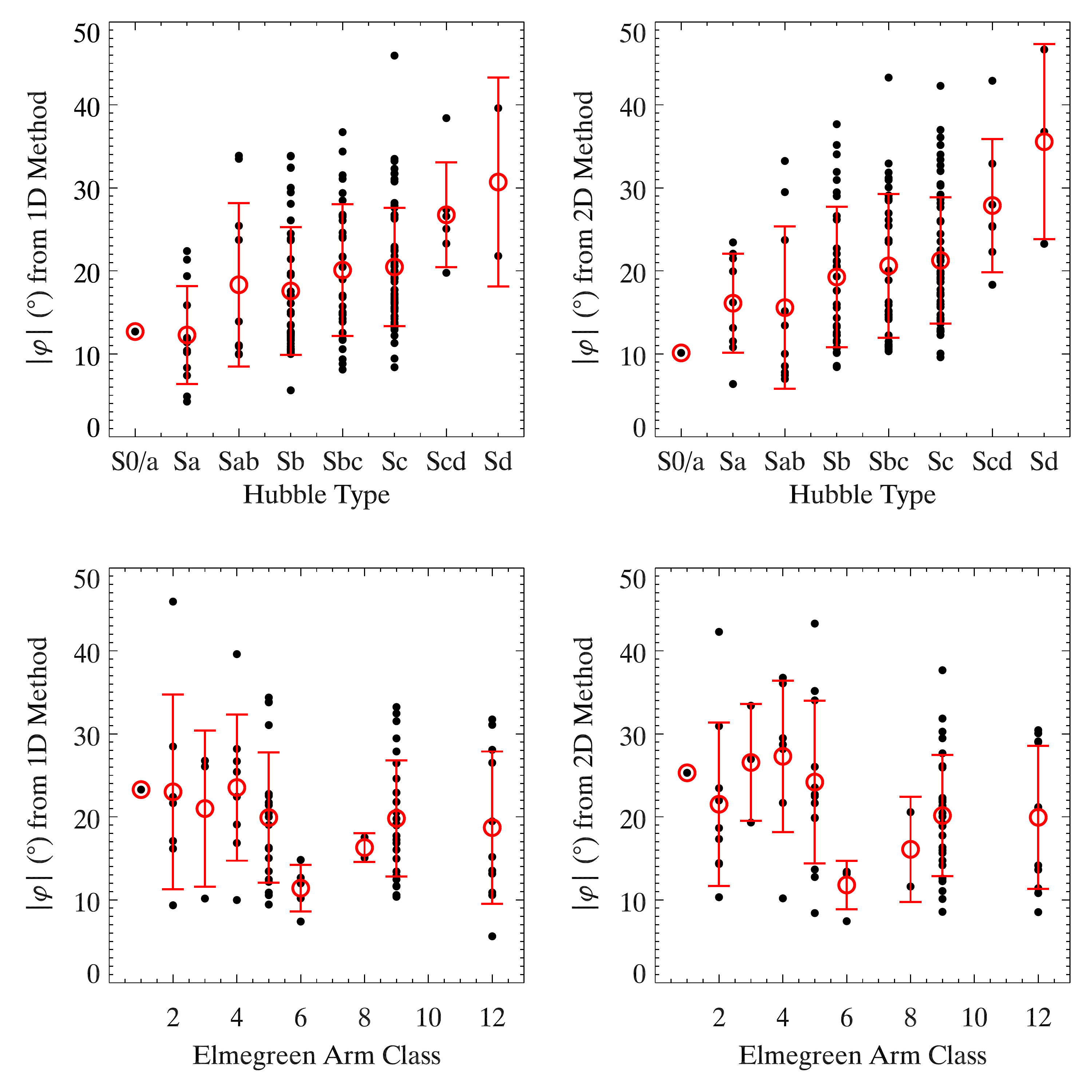}
\caption{
Dependence of pitch angle (left: 1D method; right: 2D method) on (top) Hubble type and (bottom) Elmegreen arm class.  Spiral arm pitch angle decreases with earlier Hubble type, but with large scatter ($\sim8\degr$).  Pitch angle does not correlate with Elmegreen arm class.}
\end{figure}

\subsection{Comparison with Previous Studies}

\begin{figure}
\figurenum{10}
\plotone{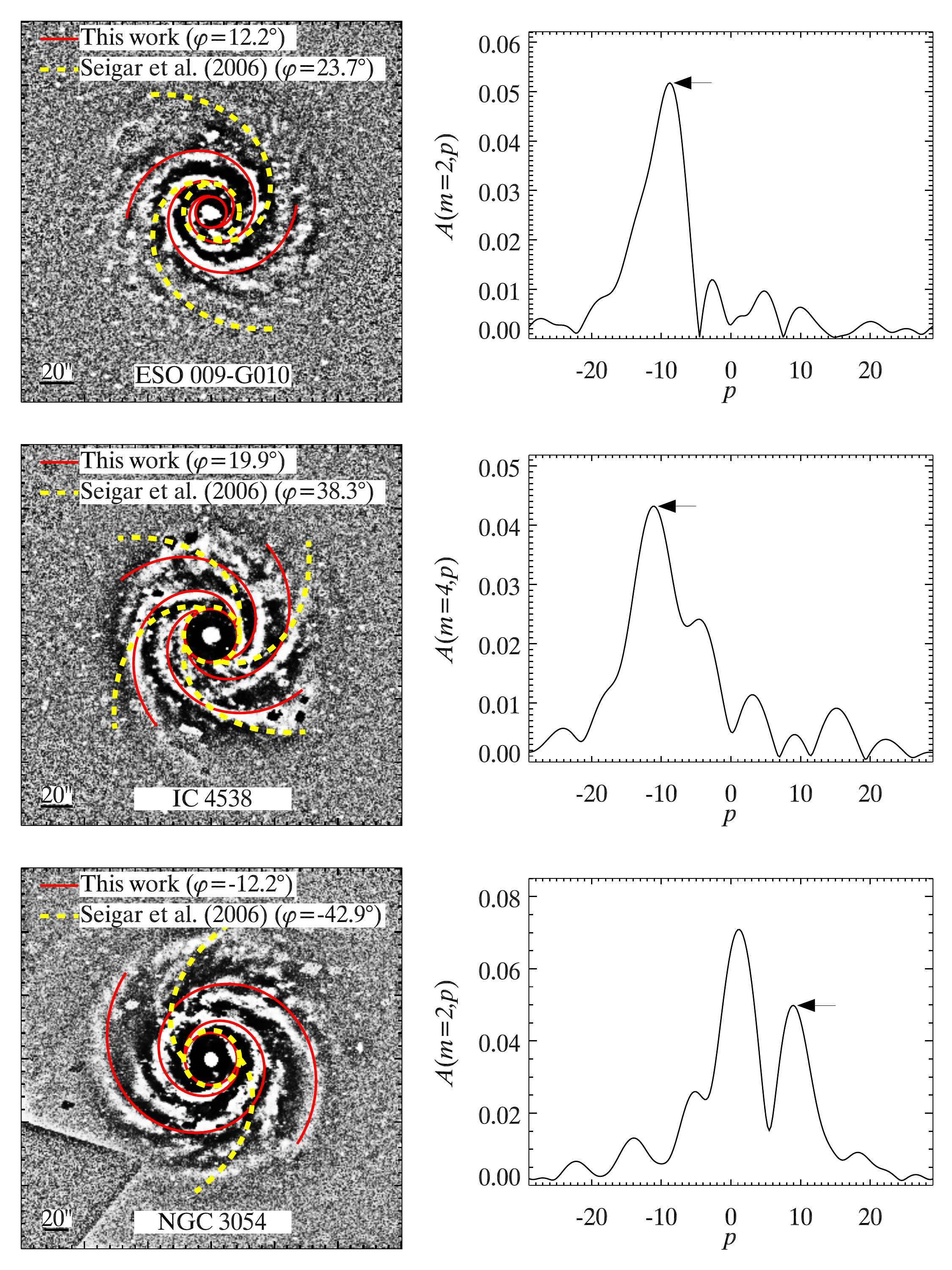}
\caption{Illustration of our new pitch angle measurements for three galaxies in common with the study 
of \cite{Seigar2006}.  (Left) Unsharp-masked image overplotted with synthetic arms with our pitch angle (solid 
red curve) and that of Seigar et al. (2006; yellow dashed curve).  The circles nearby the 
galactic center represent the innermost radial boundary. (Right) Fourier spectra 
for one of the four radial bins used to calculate the pitch angle, 
with an arrow indicating the peak chosen.}
\end{figure}

\begin{figure}
\figurenum{11}
\centering
\includegraphics[width=12cm]{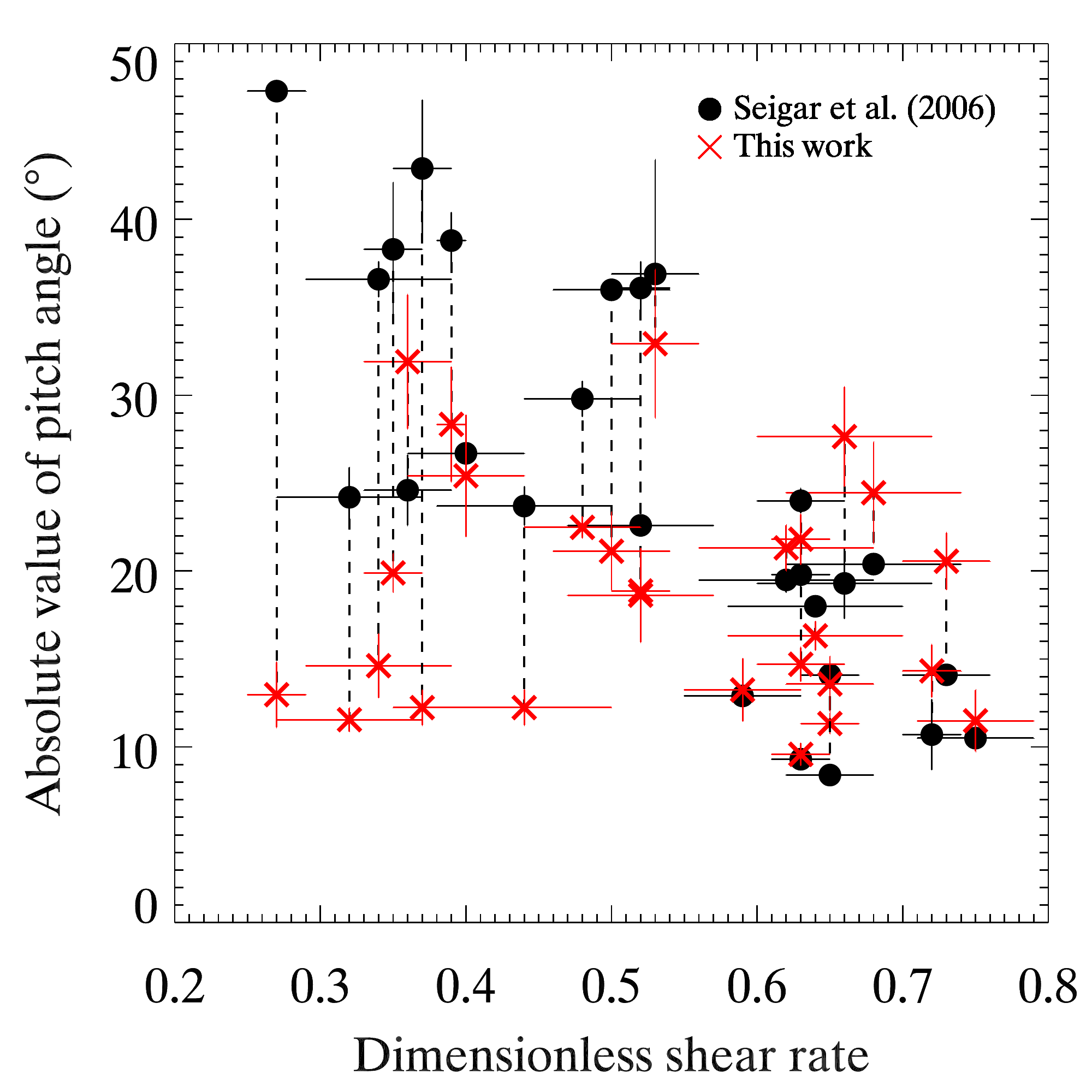}
\caption{Correlation between shear rate and pitch angle. Black filled points 
are the results from \cite{Seigar2006}, while red crosses represent 
our new measurements of pitch angle. Data points for the same galaxy are 
connected with a dashed line for comparison. Our new measurements show 
that the pitch angle correlates weakly, if at all, with the shear rate.}
\end{figure}

\cite{Seigar2005, Seigar2006} report that the pitch angle of spiral arms strongly correlates with the 
shape of the galactic rotation curve, such that
galaxies with open arms have rising rotation curves while those with
tightly wound arms have falling rotation curves. The shape of 
the rotation curve is quantified by the dimensionless shear rate, 
defined by \cite{Seigar2005} as $\frac{1}{2}(1-\frac{R}{V}\frac{\text{d}V}{\text{d}R})$, 
where {\it V} and {\it R} are the local rotation velocity and radial distance from 
the galactic center, respectively. By contrast, the sample of 13 galaxies analyzed by \cite{Kendall2015} 
does not reveal a strong correlation between arm pitch angle and shear rate.
As \cite{Seigar2006} make use of galaxy images from CGS, we can independently reexamine their results 
using our independent measurements of arm pitch angles.

\cite{Seigar2006} use 2DDFT to measure pitch angles for 31 CGS galaxies.  Among these,
10 are not in our main sample because they do not satisfy our sample selection criteria.  We uniformly 
analyze all the objects in the \cite{Seigar2006} sample, 
using, for consistency, the 2D method.
We successfully measure pitch angles for 27 of the 31 galaxies. 
The spiral arm pitch angles derived by \cite{Seigar2006} are strongly overestimated (more than 10$\degr$) for 
nine galaxies. 
For example, in the case of NGC~1566, \cite{Seigar2006} quote a pitch angle of $36.0\pm0.3\degr$, 
whereas we find  $21.1\pm2.2\degr$, which is consistent with the value of $22\pm2\degr$ given by 
\cite{Kennicutt1981} and $22.3\pm0.1\degr$ quoted by \cite{Kendall2011}.  
For this galaxy, the overestimation of pitch angle 
by \cite{Seigar2006} is caused by improper projection parameters. The projection parameters we adopt 
($e=0.21$, $\text{PA}=-2\degr$) agree with those given by Kendall et al. (2011; $e=0.23$, 
$\text{PA}=3\degr$), which were obtained by 2D bulge-to-disk decomposition using {\tt GALFIT}. 
The discrepancy for the other eight cases, however, cannot be attributed to differences in adopted 
projection parameters, which are quite similar to ours. 
We verified that adopting the same projection parameters used by \cite{Seigar2006}
does not resolve the discrepancy in pitch angles.  As Figure 10 illustrates, 
the synthetic arms created using our pitch angles trace the spiral arms very well, and 
the galaxy images illustrate that the pitch angle from \cite{Seigar2006} are severely overestimated.
Using our new pitch angle measurements of 27 galaxies and the shear rates given in
\cite{Seigar2006}, we redraw the scatter diagram for shear rate 
and pitch angle (Figure 11).  The original data of Seigar et al. (2006; their Table~3) 
are shown for comparison\footnote{The reconstructed correlation is slightly different from Figure 3 of 
Seigar et al. (2006) because three of the points in their figure are actually inconsistent 
with the data listed in their own Table~3.}.
While our new measurements exhibit a weak trend between
pitch angle and shear rate, they do not support the strong 
correlation reported by \cite{Seigar2006}. Most of the
large pitch angles ($>35\degr$) reported by \cite{Seigar2006} were
severely overestimated. Our results are 
quite similar to those of Kendall et al. (2015; their Figure 15).

Compared with previous studies, our pitch angles are more accurate, owing to 
careful determination of the radial range over which the measurements are made and, especially, 
the proper peak of the most relevant Fourier mode that 
represents the spiral arms. Our measurements lay the foundation for further quantitative studies on the
dependence of spiral arms properties on galaxy properties.

\subsection{Grand-design Spiral Arms} 

\begin{figure}
\figurenum{12}
\centering
\includegraphics[width=12cm]{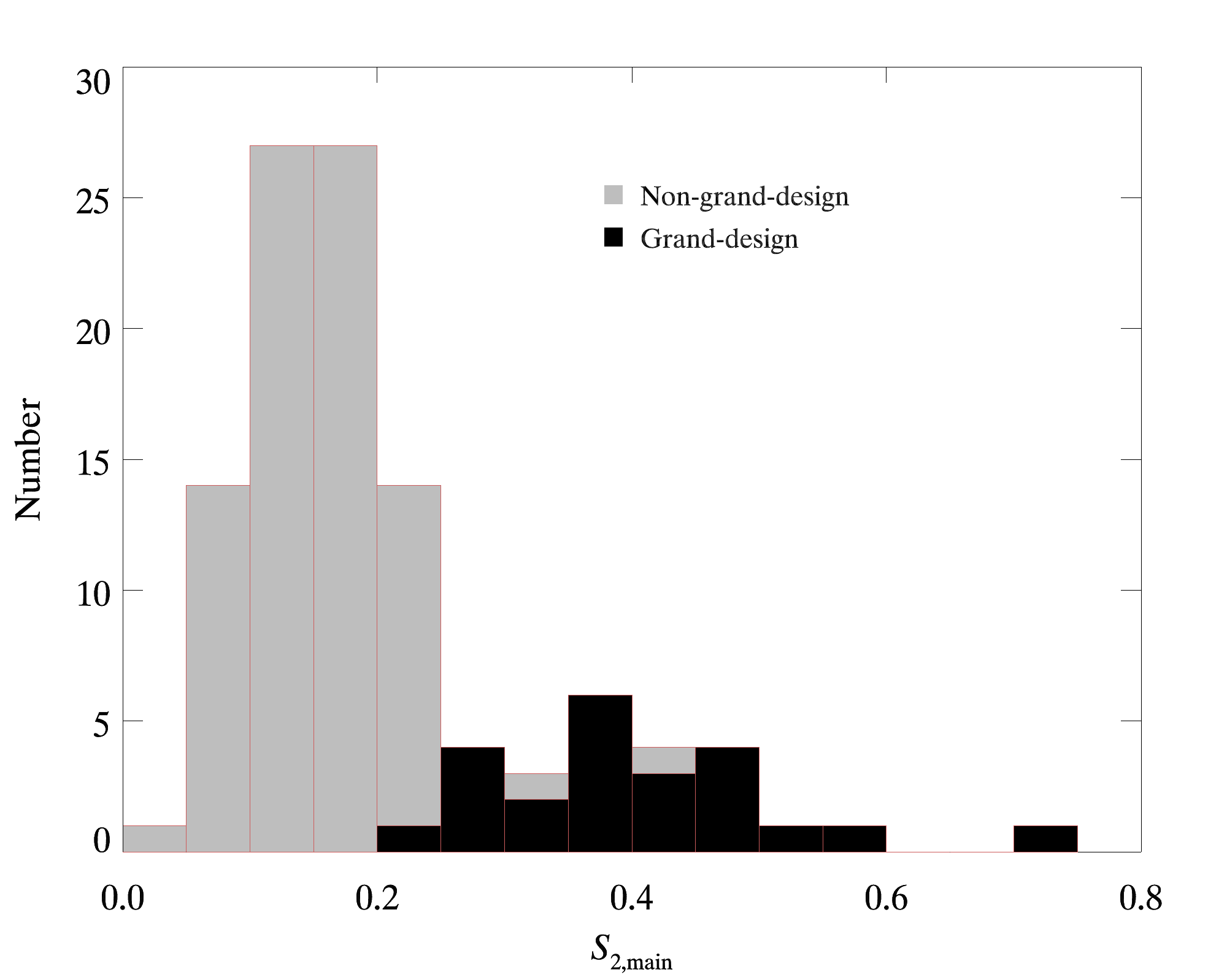}
\caption{Distribution of mean relative amplitude of the $m=2$ mode in the main 
spiral region, $S_{2,\text{main}}$, for galaxies with a dominant $m=2$ mode.
Results for grand-design galaxies are shown in the black histogram, while 
results for non-grand-design galaxies are shown in the grey histogram. 
$S_{2,\text{main}}$ is an effective, quantitative parameter to select 
grand-design galaxies.}
\end{figure}

Grand-design galaxies, as defined by \cite{Elmegreen1987, Elmegreen1995}, have 
two symmetric, long spiral arms dominating the galactic disk.  However, some 
non-grand-design galaxies can also have two symmetric arms of sufficient 
prominence that can be considered grand-design, even if the spiral arms may 
not extend to the very outer parts of the disk\footnote{The 
classification of grand-design galaxies varies from study to study.  In the 
classification system of \cite{Elmegreen1987}, grand-design galaxies are 
assigned arm class (AC) 5--12, AC 12 \citep{Elmegreen1995}, and AC 10--12 
\citep{Elmegreen2011}. On the other hand, some optically flocculent galaxies 
can have two arms in the near-infrared \citep{Block1994, Thornley1996, 
Thornley1997, Block1999, Elmegreen1999, Kendall2011}.}.  Grand-design spirals 
have a different formation mechanism than other arm classes. They may be arms 
dynamically triggered by the tidal force of a companion galaxy 
\citep[e.g.,][]{Dobbs2010}, or they may be wave modes obeying density wave 
theory \citep{LinShu64, Bertin1989a, Bertin1989b}. By contrast, irregular spiral structures 
may be the product of random gravitational instabilities generated by the 
gas and old stars in the disk \citep{Goldreich1965, Julian1966, Kalnajs1971}. 
Spiral structure that is marginally regular and relatively symmetric may be 
the result of a combination of different physical mechanisms. In our 
sample, only 109 of 211 galaxies have available arm class (AC) classifications 
in \cite{Elmegreen1987, Elmegreen1995}.  We reclassify our galaxies with 
two clearly symmetric arms
in {\it I}-band images as grand-design galaxies (Table~1).
Most of our grand-design galaxies have AC 9 and 12, if available, with one 
galaxy designated AC 4. Some galaxies with AC 12 (e.g., NGC~1357) are 
classified as non-grand-design galaxies because their arms are not 
sufficiently dominant.

Here, we propose a new empirical, quantitative method to identify grand-design 
spirals.  The main spiral 
region determined using the 1D method reflects the radial range where the 
grand-design spiral structure occupies.  It should be strongly dominated by 
the $m = 2$ Fourier mode. We define a new quantity, $S_{2,\text{main}}$, to 
denote the mean relative amplitude of the $m=2$ mode in this main spiral 
region for galaxies with a dominant $m = 2$ mode.  All those with dominant 
mode $m\neq2$ are non-grand-design galaxies.  Grand-design galaxies should 
have stronger $S_{2,\text{main}}$ than non-grand-design galaxies.  Figure 12 
shows the histogram of $S_{2,\text{main}}$ measured from {\it R}-band images. 
Two populations clearly emerge: those with $S_{2,\text{main}} \geq 0.25$ are  
grand-design galaxies; those with $S_{2, \text{main}} \lesssim 0.25$ are 
almost exclusively non-grand-design galaxies, although there are a few 
overlapping objects. Of the two non-grand-design galaxies that have 
$S_{2,\text{main}} > 0.25$, one is strongly lopsided, leading to an exaggerated 
$m = 2$ mode.  One of the grand-design galaxies has relatively weaker spiral 
structure and thus a low value of $S_{2,\text{main}}$.  Overall, 
$S_{2,\text{main}}$ appears to be an effective, quantitative parameter to 
select grand-design galaxies.  We utilize this criterion to reclassify all the 
spiral galaxies in CGS.  Table~1 lists the 23 cases that we consider to be 
grand-design spirals.

\begin{figure}[t]
\centering
\includegraphics[width=16cm]{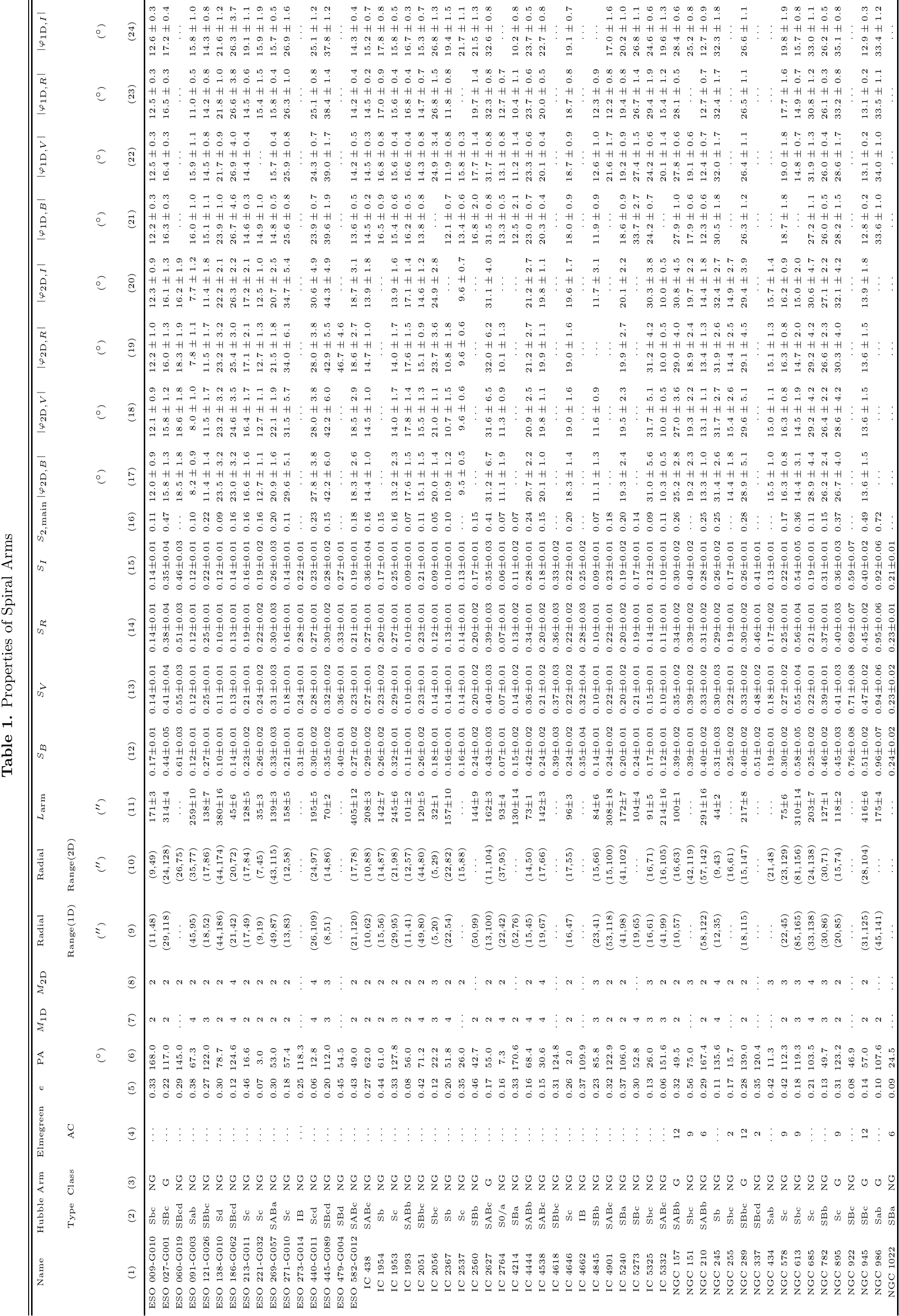}
\end{figure}
\clearpage    
\begin{figure}[t]
\centering
\includegraphics[width=16cm]{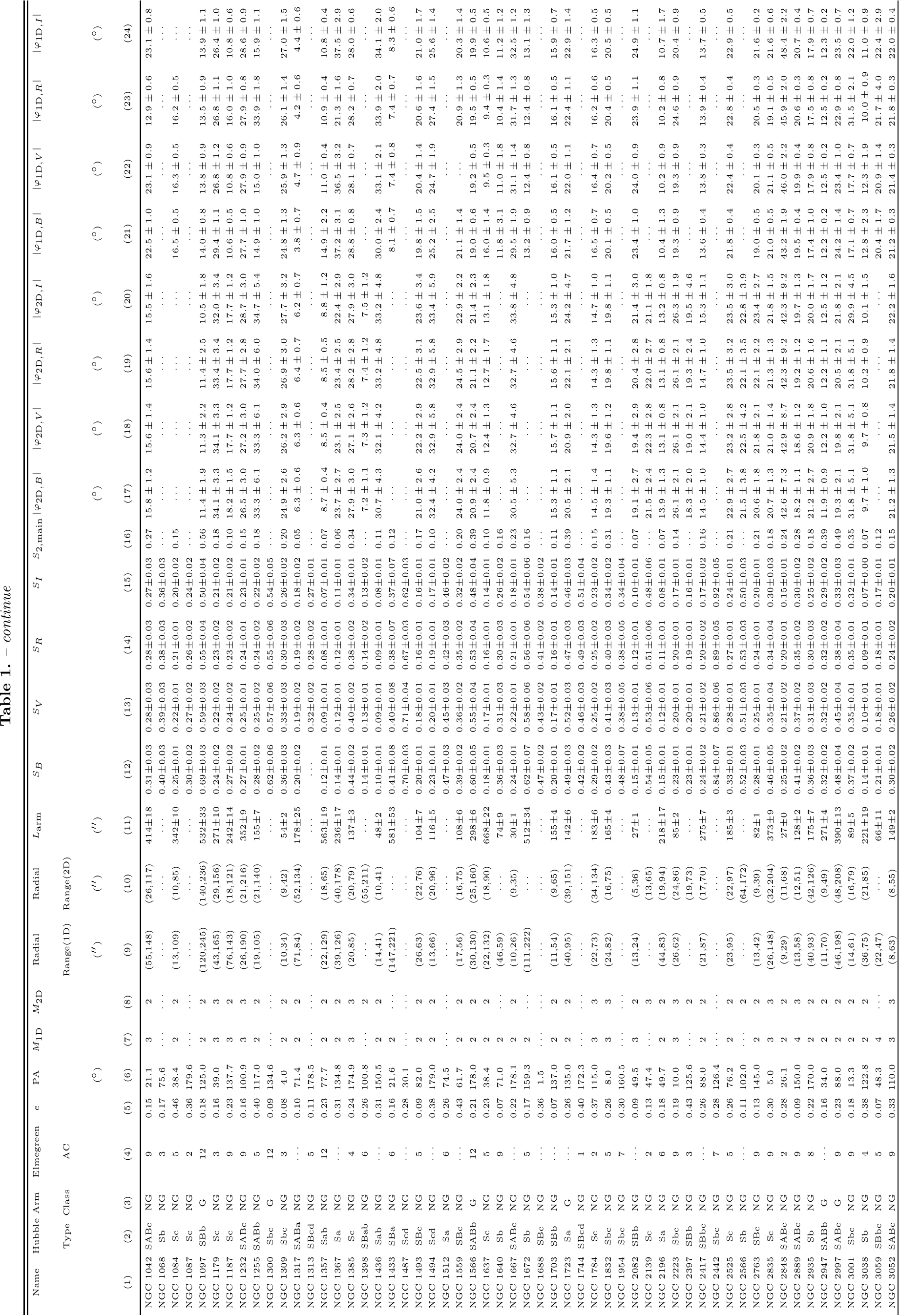}
\end{figure}
\clearpage    
\begin{figure}[t]
\centering
\includegraphics[width=16cm]{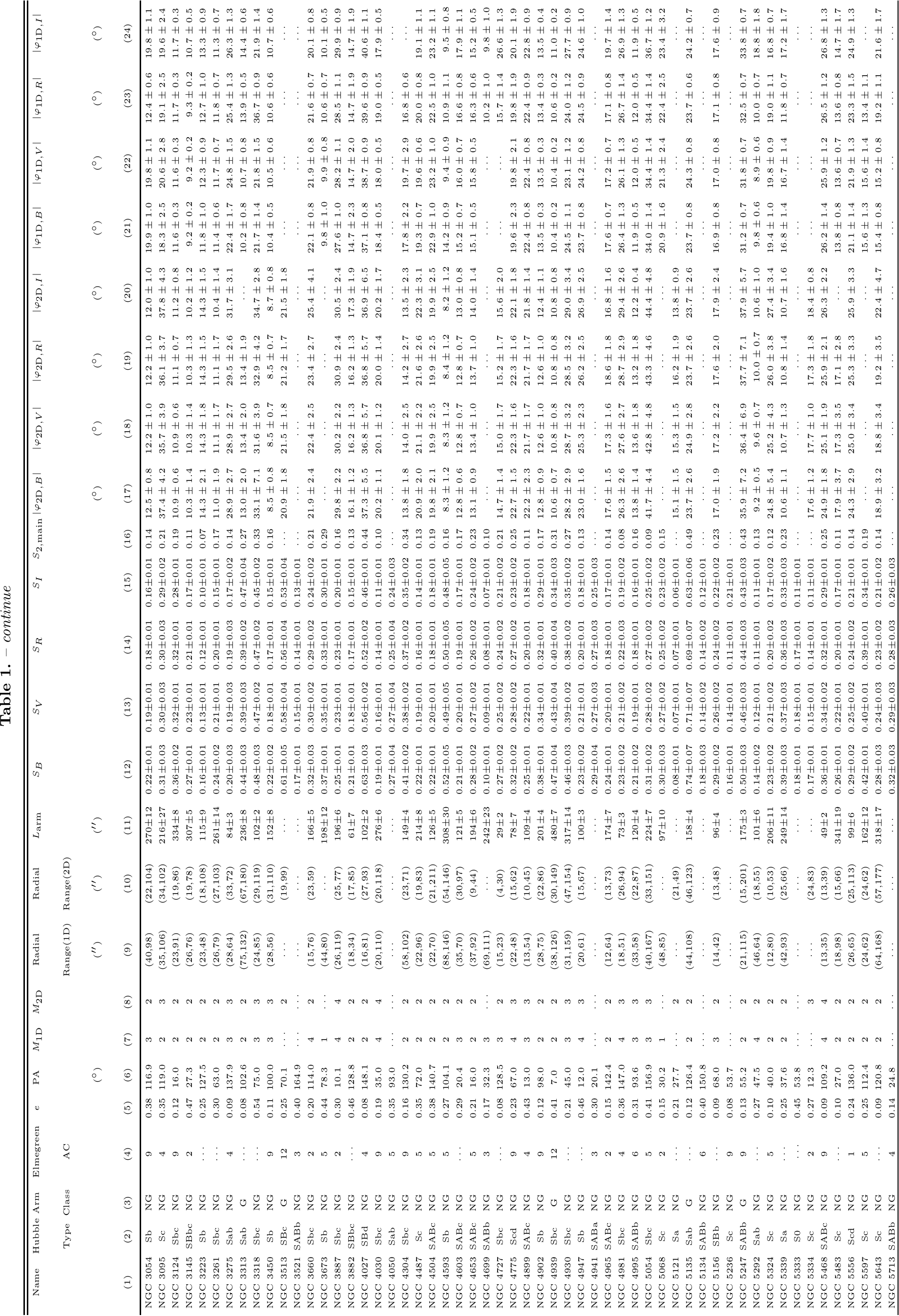}
\end{figure}
\clearpage    
\begin{figure}[t]
\centering
\includegraphics[width=14.5cm]{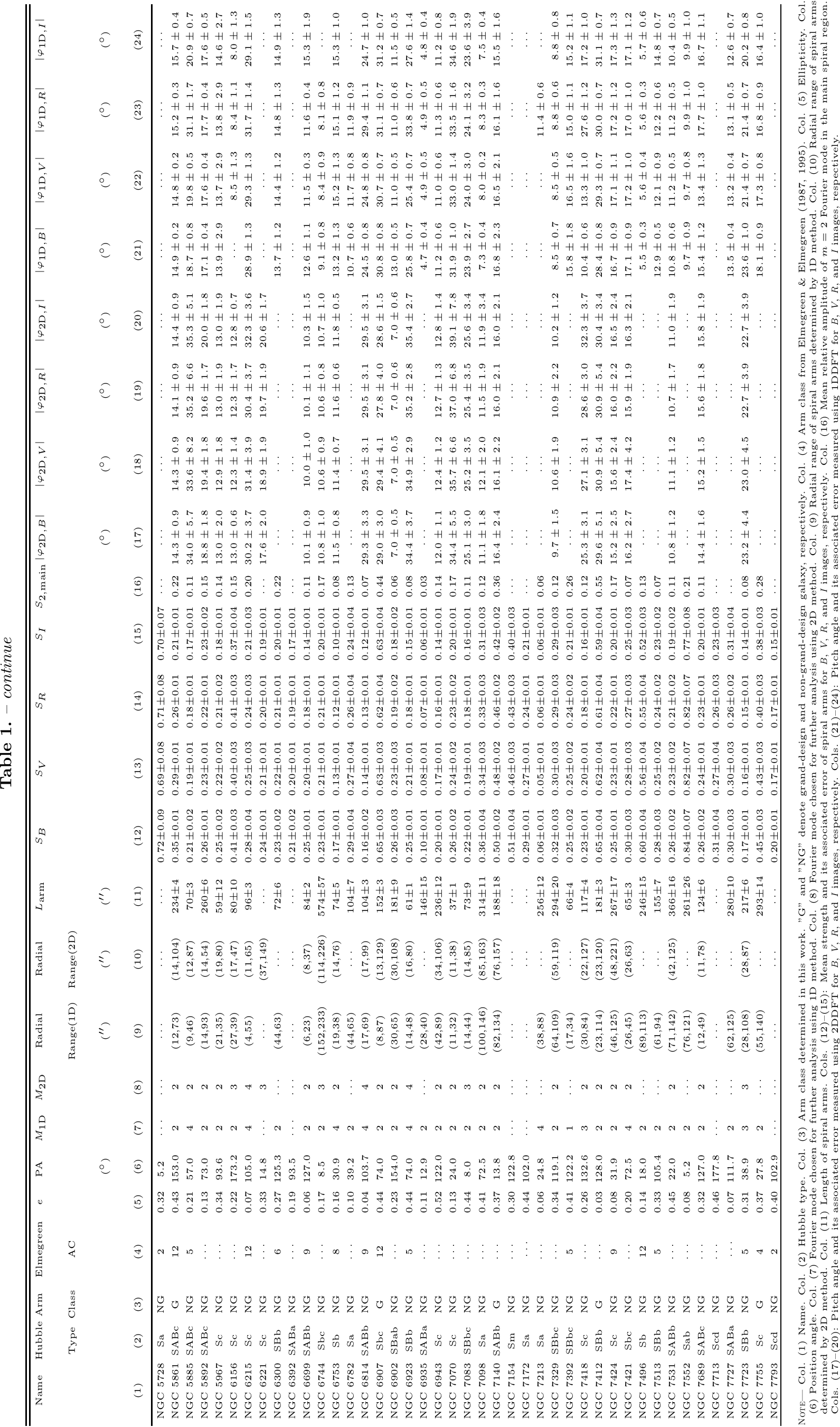}
\end{figure}
\clearpage

\section{Simulated Images and Measurement Limit} \label{sec:robust}

The availability of large wide-field galaxy surveys allows us to use 
structural parameters of spiral arms to probe the formation and evolution of 
spiral structure in a much more statistical way. For instance, we can use 
resources available from SDSS to study the possible dependence of arm strength 
or pitch angle on other global galactic properties, including stellar mass, 
bulge-to-total light ratio, and environment.  In the same spirit, we can extend 
the investigation to higher redshifts, using, for instance, images from CANDELS,
to study the cosmic evolution of spiral structure. 
A practical limitation is
that SDSS or CANDELS images may lose structural information because of 
resolution or signal-to-noise ratio (S/N) effects. Image degradation may be 
caused by instrumental or cosmological effects: exposure time,  pixel size, 
point-spread function (PSF), cosmological dimming, or cosmological angular 
size. These effects can introduce possible measurement bias or uncertainty.
In order to understand the limitations of our technique, we use high-quality 
CGS {\it R}-band images to simulate typical SDSS images and CANDELS images at 
various redshifts.  We apply to these simulated images the same 1D and 2D 
Fourier analysis described in Section 3 to see how the image quality affects
our measurements. Our analysis is similar in spirit to the
work of \cite{Block2001}, who investigated the effect of redshift on the 2D 
Fourier spectra of the spiral structure of NGC~922.

\subsection{Simulated Images}

To generate artificial SDSS and CANDELS images, we need to consider several 
observational parameters: redshift, PSF, sky noise, photometric zero point,
exposure time, and pixel size.  To simulate SDSS images, we use the parameters 
appropriate for an {\it r}-band observation, which has a pixel scale of 
$0\farcs396$, gain of 4.735, exposure time of 53.9 s, median Vega zero point 
of 23.94 mag, median PSF FWHM of $1\farcs4$, and a median sky background of 
$0.101904$ counts~s$^{-1}$.  These statistics were taken from the SDSS-DR7 
photometric catalog \citep{Abazajian2009}.  When generating the simulated 
images at higher redshift, the physical linear scale of the angular FWHM of 
the PSF and pixel size in CGS images cannot exceed that of the output simulated
images. The minimum redshift of the simulated images is set as the redshift 
where both input PSF width and pixel size begin to match the output PSF and 
pixel size.  The maximum redshift is set by the practical limitation of whether 
any useful structure can be resolved.  As Figure 13 shows, spiral arms are 
nearly smoothed out in SDSS images by $z \approx 0.1$.  We thus simulate SDSS 
images over the redshift range $0.01 \lesssim z \lesssim 0.1$, in steps of 
$\Delta z = 0.01$.
	
CANDELS targets five fields (GOODS-N, GOODS-S, UDS, EGS, and COSMOS) in two 
depths.  The shallow portion of the survey (CANDELS/Wide) has exposures in all 
five fields; the deep portion (CANDELS/Deep) focuses only on GOODS-S and 
GOODS-N \citep{Grogin2011}. We therefore generate two sets of simulated images 
for CANDELS.  The COSMOS images viewed with the Advanced Camera for Survey 
(ACS) Wide-Field Channel (WFC) detector in the F814W filter, with an exposure 
depth of 3.3 ks, are typical of CANDELS/Wide images.  We use the 
COSMOS mosaic images \citep{Koekemoer2011} to estimate the 
sky noise level of our simulated CANDELS/Wide images by averaging the sky 
noise in randomly selected sky regions. The GOODS-S deep field mosaic image
viewed with ACS WFC has exposure depth up to 31.9 ks.
To simulate the CANDELS/Deep images, 
we use the deepest portion of the GOODS-S deep field mosaic image 
\citep{Koekemoer2011} to estimate the sky noise level.  The final adopted 
parameters for simulating the CANDELS images are summarized in Table~2.

{\it HST}\ images have sufficiently high resolution that spiral structure in 
distant galaxies remains well-resolved.  However, cosmological dimming 
is a factor, as is S/N.  
We simulate CANDELS images for a redshift range from $z = 0.1$ to $z = 1.1$,
in steps of $\Delta z = 0.1$.  Starting with CGS images, we reduce their 
angular size, surface brightness, and resolution, adding random Poisson noise 
to mimic the properties of images observed at various redshifts by different 
instruments.  At $z \approx 0.1$, the {\it R} band shifts into the {\it V} band, and at 
$z \approx 1.1$ the {\it I} band roughly maps into the {\it B} band. We do not 
consider {\it k}-correction.   
Empirical PSFs of CGS images are from \cite{Ho2011}, while the PSF for the 
simulated SDSS images is approximated using a 2D Gaussian function 
with FWHM = $1\farcs4$ and for simulated CANDELS images a real
PSF with FWHM = $0\farcs1$ is used.

The Poisson sky noise of CGS images is derived from the sky level calculated 
by \cite{Li2011}.  Our simulation procedure is summarized as follows:

\begin{deluxetable}{ccccccccc}
\tablenum{2}
\tablecaption{Parameters for the  Simulated SDSS and CANDELS Images}
\tablehead{
\colhead{Survey} &
\colhead{Instrument} &
\colhead{Band} &
\colhead{Zeropoint} &
\colhead{Exposure time} &
\colhead{Sky noise} &
\colhead{Pixel size} &
\colhead{PSF FWHM} &
\colhead{Gain} \\
\colhead{} &
\colhead{} &
\colhead{} &
\colhead{} &
\colhead{(s)} &
\colhead{(counts/s)} &
\colhead{($\arcsec$)} &
\colhead{($\arcsec$)} &
\colhead{} \\
\colhead{(1)} &
\colhead{(2)} &
\colhead{(3)} &
\colhead{(4)} &
\colhead{(5)} &
\colhead{(6)} &
\colhead{(7)} &
\colhead{(8)} &
\colhead{(9)} 
}
\startdata
		SDSS & --- &{\it r} & \hd 23.94 & 53.9 &                           \hb \hc \hmin \hmin 0.101904    &    0.396 & 1.4   & 4.735\\
		CANDELS/Wide & ACS/WFC & F814W & 25.526 & 6900 & 0.00231366  &   \hd0.03 &   0.1 & 1 \\
		CANDELS/Deep & ACS/WFC & F814W & 25.526 & 32000 & 0.00107436  & \hd0.03 &   0.1 & 1 \\
\enddata
\end{deluxetable}

\begin{enumerate}
\item We rebin the CGS image and its associated PSF by taking into account the 
effect of pixel size and cosmological reduction of galaxy angular size. The 
rebinning factor is

\begin{eqnarray}
F_{\text{rb}} = \frac{n_0}{n^{\prime}} = \frac{d^{\prime} / (1+z^{\prime})^2}{d_0 / (1+z_0)^2} \cdot \frac{p^{\prime}}{p_0},
\end{eqnarray}

\noindent
where $n$, $d$, $z$, and $p$ represent the total pixel number, luminosity 
distance, redshift, and pixel size, respectively. The luminosity distance is 
calculated assuming $\Omega_m = 0.3$ and $\Omega_{\lambda} = 0.7$.  The 
subscript 0 and prime superscript denote CGS parameters and 
corresponding simulation parameters, respectively.

\item We rescale the image flux with a rescale factor 

\begin{eqnarray}
F_{\text{rescl}} = \left(\frac{d^{\prime}}{d_0} \right)^2 \cdot 10^{0.4\cdot (z^{\prime}-z_0)}.
\end{eqnarray}

\noindent
Here we consider the bolometric surface brightness dimming caused by the 
redshift effect and the cosmological evolution of surface brightness reported 
in \cite{Barden2005}, who found that the surface brightnesses of galaxies at 
$z \approx 1$ are 1 magnitude brighter than those of galaxies in the local 
Universe.	

\item We convolve the resulting images from step (2) with a Gaussian kernel to 
reach the desired output PSF. The kernel is calculated using Fourier 
deconvolution of the target PSF with the rebinned CGS PSF:

\begin{eqnarray}
{\rm kernel} = \mathcal{F}^{-1} \left[ \frac{ \mathcal{F}\left({\rm PSF}_{\text{out}} \right)}{ \mathcal{F}({\rm PSF}_{\text{in}})} \right],
\end{eqnarray}

\noindent
where $\mathcal{F}$ and $\mathcal{F}^{-1}$ represent Fourier transformation and 
inverse Fourier transformation, respectively. No convolution is applied if the 
two PSFs are comparable.
	
\item For simulating CANDELS images, the Poisson noise from the original CGS 
images is negligible after flux rescaling and PSF convolution, but it may 
become considerable in the case of simulating low-redshift SDSS images. 
Thus, the noise from the original CGS images is subtracted in quadrature from 
the target noise level, including sky noise and galactic flux noise, and then 
the resulting noise map is added to the images from step (3). We use the IRAF 
task {\tt mknoise} to generated Poisson noise.
\end{enumerate}

\begin{figure}[ht!]
\figurenum{13}
\plotone{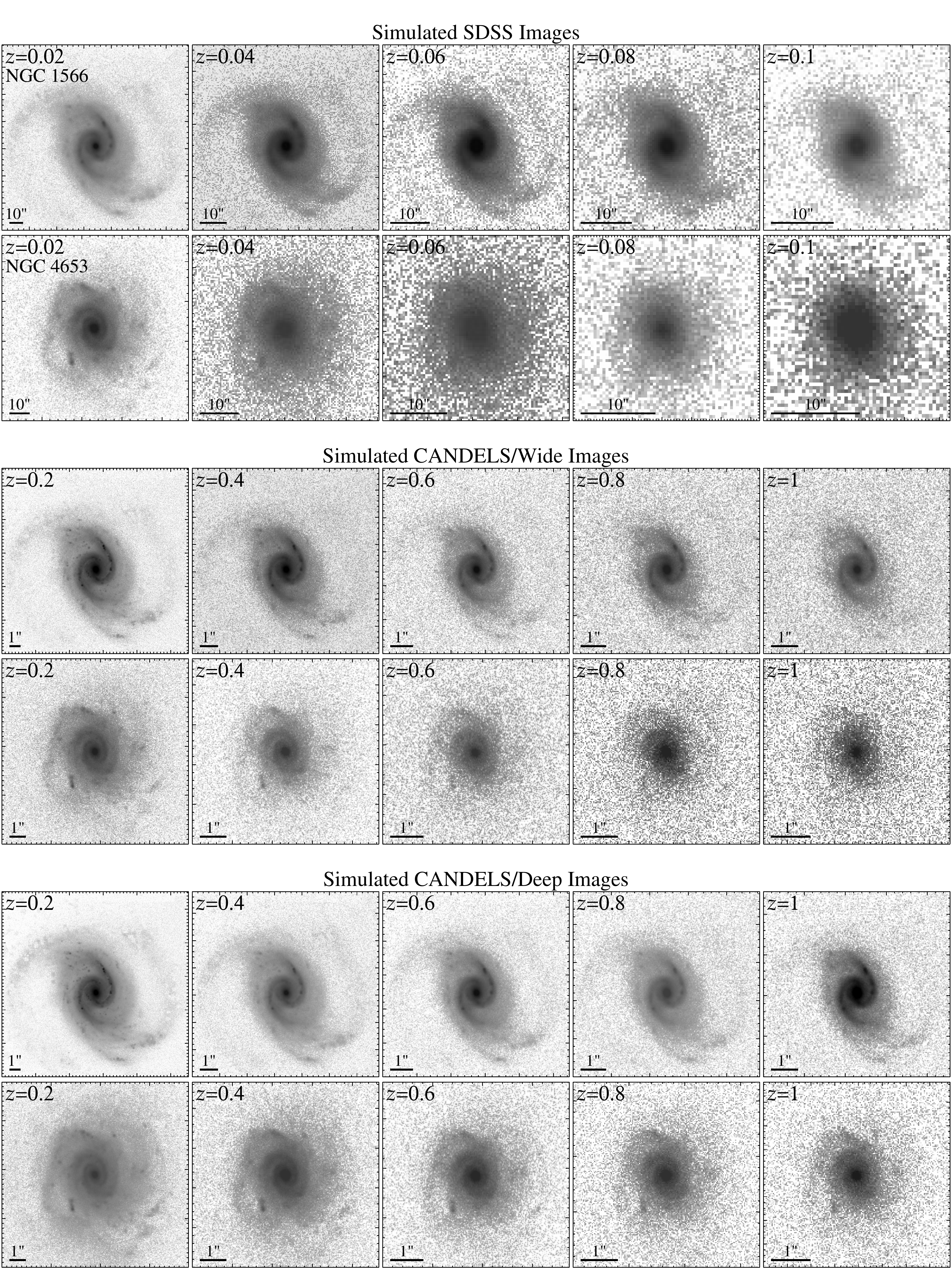}
\caption{Nearby CGS galaxies NGC~1566 (grand-design) and NGC~4653 
(non-grand-design) simulated to resemble observations by (top two rows) SDSS 
at $z$ = 0.02, 0.04, 0.06, and 0.1 and the {\it HST}\ CANDELS (middle two 
rows) Wide and (bottom two rows) Deep surveys at $z$ = 0.2, 0.4, 0.6, 0.8, and 
1.} 
\end{figure}

In Figure 13, the upper two rows present simulated SDSS images of the CGS 
galaxies NGC~1566 (grand-design spiral) and NGC~4653 (non-grand-design spiral),
as viewed at $z = 0.02$, 0.04, 0.06, and 0.1.  The middle and bottom two rows 
pertain to simulated images under conditions similar to the CANDELS/Wide and 
CANDELS/Deep fields at $z = 0.2$, 0.4, 0.6, 0.7, 0.8, and 1, respectively. For 
the simulated SDSS images, the two symmetric, strong arms of the grand-design 
galaxy can still be discerned at $z=0.1$, although they become quite blurred. 
The relatively weaker spiral structure of the non-grand-design galaxy is 
almost completely smoothed out by $z$ \gax\ 0.06.  By contrast, the resolution 
of the simulated {\it HST}\ images is good enough to recognize the inner part 
of the grand-design spiral structure in NGC~1566 out to at least $z \approx 1$, 
because this galaxy is intrinsically bright and the contrast between the arm 
and inter-arm region is high, while the structure information of 
non-grand-design spiral structure in NGC~4653 is nearly washed out by the 
noise at $z$ \gax\ 0.6 in the CANDELS/Wide field.  The CANDELS/Deep 
images obviously have better S/N and hence can go deeper.

\subsection{Robustness of Measurements for the Simulated Images} \label{sec:simulation}

The mean strength and pitch angle of spiral arms are the two most important 
parameters that can provide clues about the formation mechanism or evolution 
of spiral structure. Prior to any statistical study using SDSS or CANDELS 
images, an important step is to understand the robustness of the measurements 
under conditions that closely mimic those of the actual observations.

\subsubsection{Robustness of Mean Arm Strength Measurement}
    
Keeping $e$, PA, and bar radius as before for its CGS 
counterpart, we measure the mean strength of spiral arms following the same 
procedure described in Section 3 to extract the azimuthal light profile of 
isophotes.  We then perform 1D Fourier decomposition to calculate the relative 
amplitudes of the Fourier modes, the phase angle profile of the $m=2$ Fourier 
mode, and $A_{\rm tot}(r)$.  Figure 14 shows the results based on the 
grand-design galaxy NGC 1566, while Figure 15 displays the results for the 
non-grand-design galaxy NGC~4653, separately highlighting conditions 
appropriate for 
SDSS (left column), CANDELS/Wide (middle column), and CANDELS/Deep (right 
column).  For the simulated SDSS images, the $A_{\rm tot}(r)$ profile gradually
reduces in amplitude and becomes flattened with increasing redshift due to the
smoothing effects of the PSF (Figure 13). The mean arm strength is therefore 
systematically underestimated.  The behavior of the simulated {\it HST}\ images 
depends greatly on the intrinsic properties of the arms.  In the case of 
grand-design spiral, its clear, distintive arms remain very well detected 
without any noticeable bias out to $z \approx 1$, both in the CANDELS Wide and 
Deep fields (Figure 14).   The non-grand-design case fares far worse.  Its 
lower contrast arms get lost in the noise for $z$ \gax\ 0.6, beyond which
$A_{\rm tot}(r)$ becomes divergent and the mean strength is severely 
overestimated (Figure 15).  

To quantify the systematic bias in arm strength measurement, we calculate the 
difference between mean arm strength from the simulated SDSS/CANDELS images 
and that originally measured from the CGS images, for the full sample of 211 
CGS galaxies. 
Figure 16 shows the difference of mean arm strength as a function of redshift.
For every redshift bin, we obtain the mean value and the standard deviation as 
the measurement bias and uncertainty. The detailed values are listed in Table~3.  The 
mean arm strength is reproducible within a reasonable scatter for the 
simulated SDSS and CANDELS images. 

We note, in passing, that our method is of particular relevance for
studies of the morphological trasformation of galaxies in galaxy clusters. 
Since the work of \cite{Butcher1978, Butcher1984}, it has been known that 
galaxy clusters as recent as $z \approx 0.4$ contain a much larger 
fraction of blue, mostly spiral galaxies than nearby clusters of similar 
richness and compactness, implying a strong morphological transformation from 
spirals to S0s (e.g., Dressler et al. 1994; Fasano et al. 2000).  It would be 
of interest to apply our quantitative method in the context of these 
morphological studies, properly taking into account possible biases introduced
by the observational effects investigated here.

\begin{figure}[ht!]
\figurenum{14}
\plotone{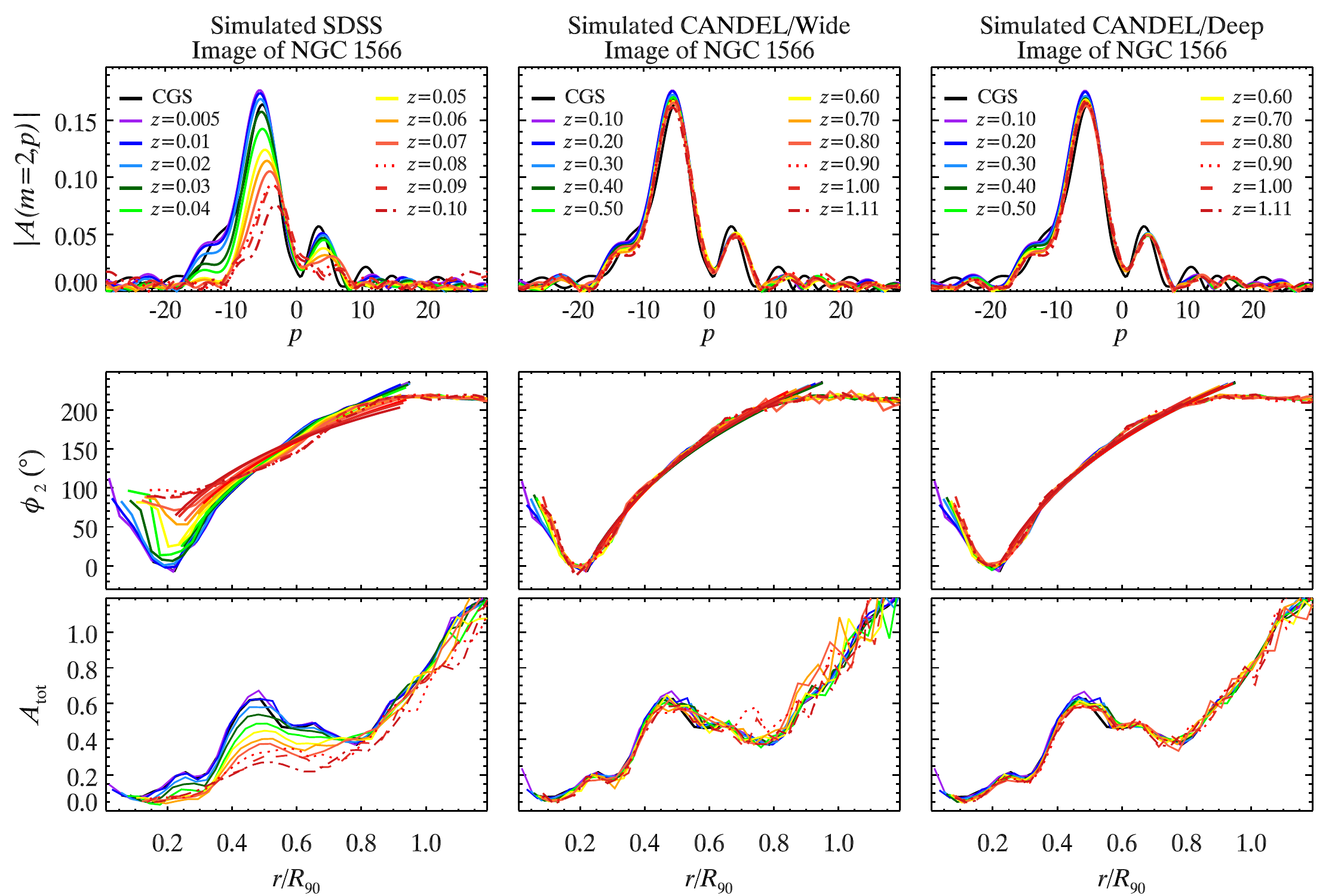}
\caption{Spiral arm properties of NGC~1566 as measured from simulated images 
observed with (left) SDSS, (middle) CANDELS/Wide), and (right)
CANDELS/Deep, for the redshifts indicated in the legends.  The 
measured quantities are (top) the 2D Fourier spectra and radial profiles
(normalized to $R_{90}$) of (middle) the phase angle of the $m=2$ Fourier 
mode and (bottom) the arm strength $A_{\rm tot}$.}
\end{figure}

\begin{figure}[ht!]
\figurenum{15}
\plotone{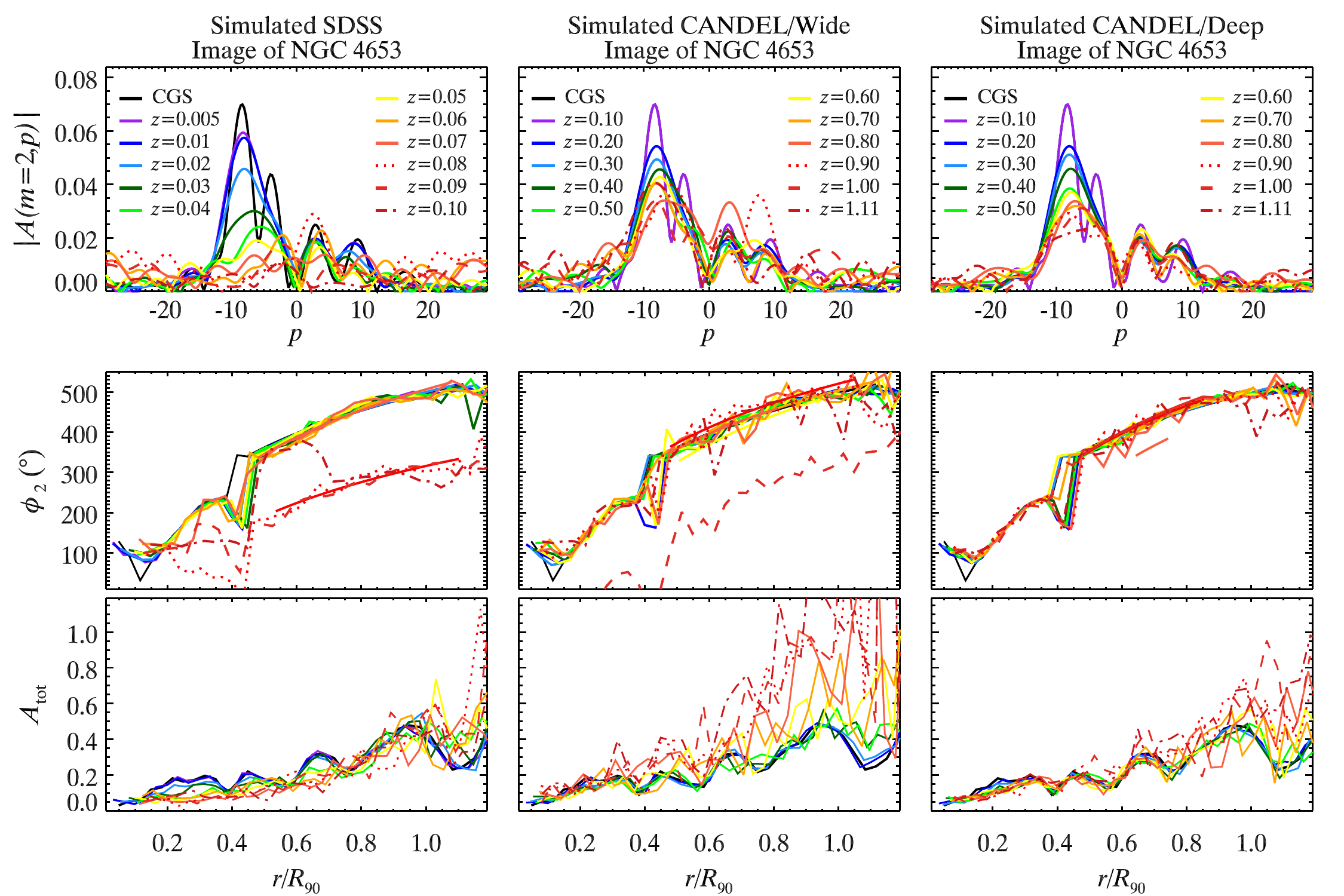}
\caption{Spiral arm properties of NGC~4563 as measured from simulated images.
Same convention as Figure 14.}
\end{figure}

\begin{figure}[ht!]
\figurenum{16}
\plotone{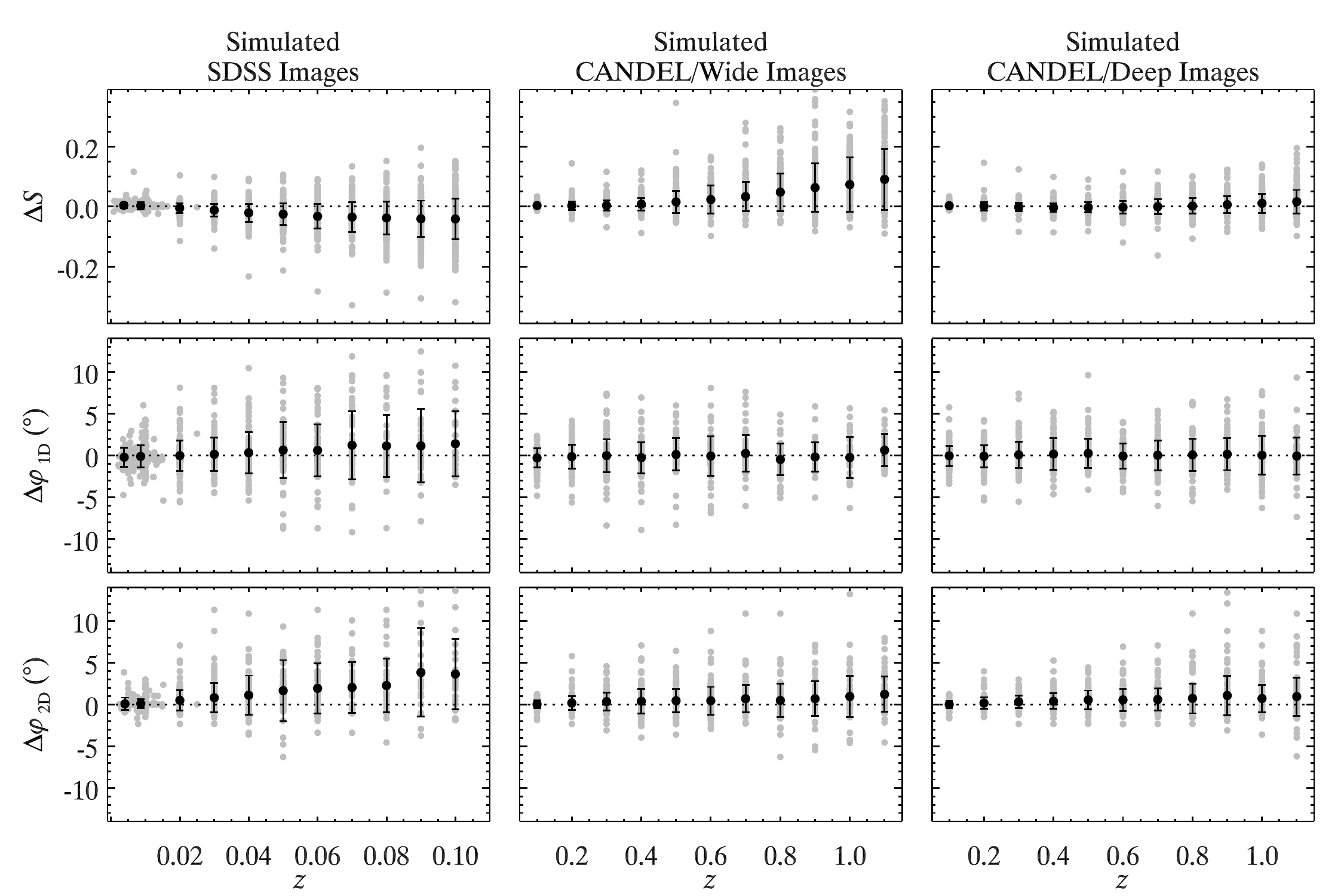}
\caption{Difference between (top) the mean strength of spiral arms,
(middle) 1D pitch angle, and (bottom) 2D pitch angle as 
measured in simulated images and observed images from CGS, as a function of 
redshift.  The left, middle, and right columns simulate conditions appropriate 
for the SDSS, CANDELS/Wide, and CANDELS/Deep surveys, respectively.
Solid points show the mean difference in each redshift bin, and the vertical 
bars represent the $1 \sigma$ standard deviation.}
\end{figure}
     
\subsubsection{Robustness of Pitch Angle Measurement}

Lastly, we examine the robustness of pitch angle measurements to effects of 
resolution, noise, and redshift.  We measure the pitch angle using both the 
1D and the 2D techniques, following exactly the procedures outlined in Sections
3.2 and 3.3, respectively.  For the 1D analysis, because of the degradation of 
image quality, the phase angle profile may not be the same as that of 
its CGS counterpart, and hence the main spiral region is redetermined according
to the behavior of the phase angle profile. For the 2D analysis, we adopt the 
same deprojection, $e$, and PA as the CGS images, and we rescale the inner and 
outer boundary of the spiral structure of the CGS images by the rebinning 
factor.  

The projection parameters of high-redshift galaxies can be readily 
determined using {\tt GALFIT}, at least to $z \approx 2$. \cite{Davari2016} 
simulated two-component model galaxies to test the limits of measuring the 
structural components of massive galaxies, finding that the disk component 
can be measured with little difficulty even at $z=2$. Since 2D Fourier spectra 
are insensitive to the exact outer radius, the outer radius can be set simply 
as the radius where the spiral arms almost disappear. A more important source 
of uncertainty for high-redshift galaxies is the determination of the inner 
radius, especially in the presence of a bar.  Fortunately, CANDELS images at 
$z$ \lax\ 1 have sufficiently high resolution and depth that bars usually can 
still be discerned.

As image quality (especially spatial resolution) deteriorates, the 
spiral structure gets washed out by the PSF or noise.  The 2D spectra have 
gradually less pronounced peaks and tend to become symmetric about $p=0$.  The 
1D phase angle profile increases in scatter.  For sufficiently low 
resolution or S/N, the 1D phase angle profile and 2D Fourier spectra can 
become completely uninterpretable.  These images were excluded. As expected, 
the number of galaxies for which pitch angle can be measured successfully 
decreases with increasing redshift (Table~3).  For simulated SDSS images at 
$z=0.1$, pitch angles can be measured for only $\sim$~20\% of the original 
sample.  Owing to the high resolution and depth of the {\it HST}\ observations,
the success rate rises to $\sim$~50\% for CANDELS/Wide and $\sim$~70\% for 
CANDELS/Deep at $z=1.1$.

We give examples of the 1D phase angle profile of the $m=2$ Fourier mode and 
2D Fourier spectra for the simulated images.  In the case of the SDSS images of
the grand-design galaxy NGC~1566 (Figure 14), its phase angle profile 
flattens as redshift increases, leading to systematic overestimation of the 
pitch angle.  The same holds true for the Fourier spectra; the peak of the $m 
= 2$ mode gradually shifts toward zero with increasing redshift, leading to 
larger and larger pitch angles.  A similar effect was noted by 
\cite{PengTianrui2017}, who found preferentially larger pitch angles for 
simulated high-redshift galaxies.  We see no systematic effects in the 
simulated CANDELS images.  The SDSS image quality has far more adverse effects 
on the Fourier spectra than the phase angle profile of the non-grand-design 
galaxy NGC~4653 (Figure 15). Its Fourier spectra decay rapidly with 
redshift, becoming almost totally chaotic at $z$ \gax\ 0.04, consistent with 
the visual appearance of the simulated images in Figure 13.  Consequently, we 
exclude the simulated images for NGC~4653 with $z$ \gax\ 0.04 from measuring pitch angle when using the 2D method.
Even with {\it HST}-quality images, its phase angle profile gradually becomes
chaotic in the outer parts, and slight shifts in the peak of the Fourier 
spectra lead to changes of pitch angle with redshift.  Contribution from the 
$m = 2$ mode is reduced gradually, and contamination from higher values of $p$ 
becomes more and more significant with increasing redshift.  These trends are
quantified explicitly in Figure 16 and in Table~3.

\def\h{\hskip -4 mm}
\def\ha{\hskip -5 mm}
\def\hb{\hskip -2 mm}
\def\hc{\hskip -2.3 mm}
\def\hd{\hskip -1.6 mm}

\begin{figure}[t]
\centering
\includegraphics[width=9cm]{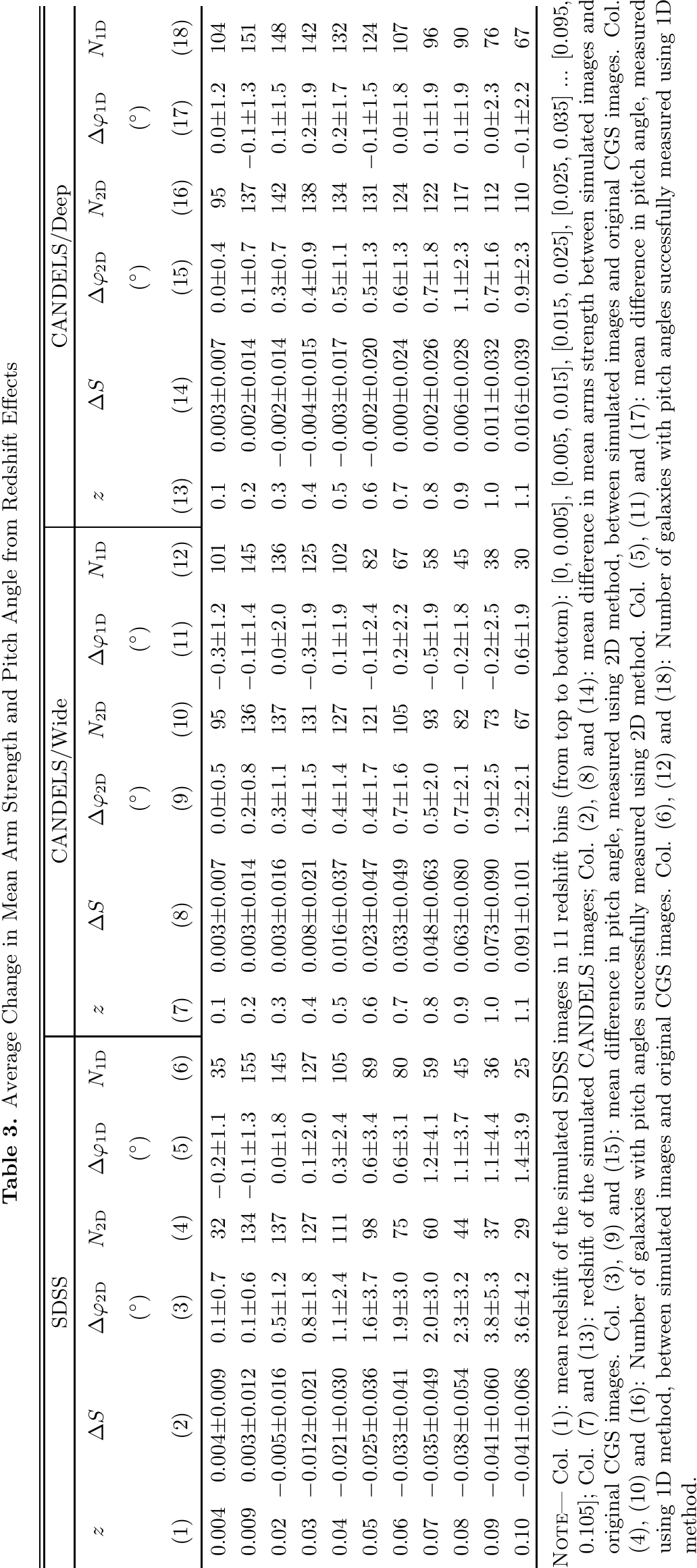}
\end{figure}
\clearpage

\section{Summary}

We use observations from CGS to develop a 
systematic method to quantify the main characteristics of galaxy spiral arms, 
including the arm number, mean arm strength, arm length, and pitch angle. The 
arm number and arm length reflect the dominant mode and continuity of the 
arms, whereas the mean arm strength reveals the relative contrast between 
spiral arms and axisymmetric disk components.  Consistent with the expectation 
that young stars form preferentially in spiral arms, the mean arm strength is 
systematically stronger toward bluer bands.  We devise an effective, new 
parameter, the relative amplitude of the $m = 2$ Fourier mode in the main 
spiral region ($S_{2, \text{main}}$), to quantitatively identify galaxies with 
grand-design spiral arms.  Grand-design spirals may owe their origin to 
external dynamical perturbation and hence may be a useful probe of the 
near-field galactic environment.  The pitch angle, which describes the 
tightness of the spiral arms, may reflect the underlying velocity field or 
mass distribution of the galaxy.  We demonstrate that consistent pitch angles 
can be derived using either Fourier decomposition of the 1D azimuthal light 
profile of isophotes or from Fourier transformation of the 2D light 
distribution.

Our methodology can be applied to measure the statistical properties of spiral 
arms in large samples of galaxies, both in the nearby 
Universe to investigate their correlation with other physical properties and 
at higher redshift to study their possible cosmological evolution. 
To this end, we use the local high-quality CGS galaxy images to generate a 
series of simulated images to mimic observing conditions typical of the 
SDSS ($z$ \lax\ 0.1) and {\it HST}/CANDELS (0.1 \lax\ $z$ \lax\ 1.1).  
We apply our analysis methods to these simulated images to understand the 
limits of their applicability and possible sources of systematic bias 
and uncertainty.  SDSS-quality images are mainly limited by their relatively 
poor angular resolution.   The mean arm strength tends to be underestimated 
and the measurement uncertainty of pitch angle reaches $\sim 5\degr$.  
{\it HST}\ images typical of CANDELS, on the other hand, are mainly restricted 
by their relativelty low S/N.  Nevertheless, both mean arm strength and arm 
pitch angle can be determined up to $z \approx 1$ without much significant bias 
or uncertainty.

\begin{acknowledgements}
We thank Iv\^{a}nio Puerari, the referee, for constructive criticisms that helped to improve the 
quality and presentation of the paper.
This work was supported by the National Key R\&D Program of China 
(2016YFA0400702) and the National Science Foundation of China (11473002, 
11721303). ZYL is also supported by a China-Chile joint grant from 
CASSACA, the Youth Innovation Promotion Association of Chinese Acadamy
of Sciences, and a LAMOST 
Fellowship, which is supported by Special Funding for Advanced Users, budgeted 
and administered by the Center for Astronomical Mega-Science, Chinese Academy 
of Sciences (CAMS).
\end{acknowledgements}

\listofchanges

\end{document}